\documentclass[9pt]{article}
\usepackage{extsizes}
\usepackage[super,sort&compress,comma]{natbib} 
\usepackage[version=3]{mhchem}
\usepackage[left=1.5cm, right=1.5cm, top=1.785cm, bottom=2.0cm]{geometry}
\usepackage{balance}
\usepackage{mathptmx}
\usepackage{sectsty}
\usepackage{graphicx} 
\usepackage{lastpage}
\usepackage[format=plain,justification=justified,singlelinecheck=false,font={stretch=1.125,small,sf},labelfont=bf,labelsep=space]{caption}
\usepackage{float}
\usepackage{fancyhdr}
\usepackage{fnpos}
\usepackage[english]{babel}
\addto{\captionsenglish}{%
  
}
\usepackage{array}
\usepackage{droidsans}
\usepackage{charter}
\usepackage[T1]{fontenc}
\usepackage[usenames,dvipsnames]{xcolor}
\usepackage{setspace}
\usepackage[compact]{titlesec}
\usepackage{hyperref}

\usepackage{stfloats} 

\definecolor{cream}{RGB}{222,217,201}

\begin{document}

\LARGE{Graphical abstract}\par\nobreak\vspace{1em}
    \includegraphics[scale=1]{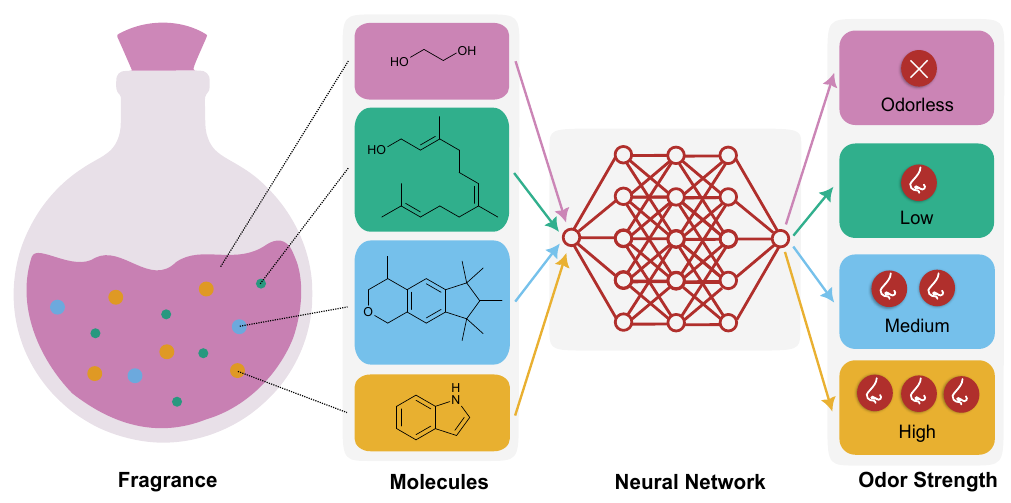}

\clearpage



\begin{center}
\LARGE{\textbf{Machine learning for smell: Ordinal odor strength prediction of molecular perfumery components}}
\end{center}

\vspace{5mm}

\begin{center}
    \large{Peter Fichtelmann\textit{$^{a}$} and Julia Westermayr\textit{$^{a,b}$}}
\end{center}

\vspace{3mm}

\begin{center}
\normalsize
    $^{a}$~Wilhelm-Ostwald Institute of Physical and Theoretical Chemistry, Leipzig University, Linn\'{e}stra{\ss}e 2, 04103 Leipzig, Germany \\
    $^{b}$ Center for Scalable Data Analytics and Artificial Intelligence Dresden/Leipzig, Humboldtstra{\ss}e 25, 04105 Leipzig, Germany
\end{center}

\vspace{10mm}

\normalsize


Predicting olfactory perception directly from molecular structure is central to fragrance design that plays a role in a wide range of industries, such as perfumery, food and beverage, and health care. Among olfactory attributes, odor strength is a key factor in shaping odor perception, but its modeling has been impeded by scarce and fragmented intensity data. In this work, we introduce an ordinal odor strength data set of over 2,000 molecules by integrating two different public sources, mapping structures to odorless, low, medium, and high categories.
Across several molecular encodings and supervised learning algorithms we compared different prediction strategies. Dimensionality reduction and SHAP analysis identifies molecular size, polarity, ring features, and branching as primary drivers, consistent with mass‑transport constraints on volatility, sorption, and receptor access.
This scalable ordinal framework enables reliable odor‑strength estimation for novel molecules and provides a foundation for in silico fragrance design.

\clearpage

\section{Introduction}

Fragrances are ubiquitous in our daily lives. We encounter them every day in a variety of products, including perfumes, cleaning agents, hygiene articles, or care needs. However, the creation of a fragrance is a tedious process that requires fine-tuning of combinations and concentrations of up to hundreds of different raw materials. Consequently, fragrance design is costly and reserved to only a few hundred highly trained individuals worldwide with years of experience known as perfumers. For comparison, more astronauts are currently alive than perfumers. \cite{wiltschko_2023}

\begin{figure}[!h]
    \centering
    \includegraphics[scale=1]{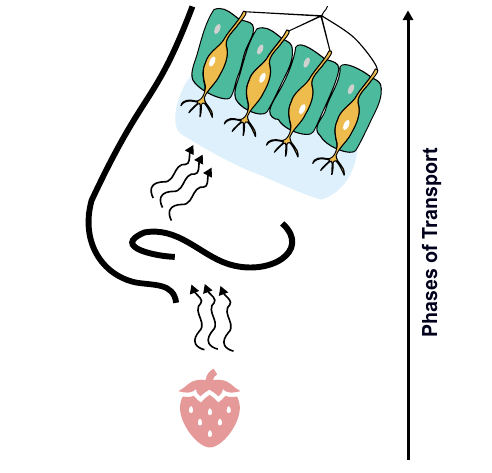}
    \caption{Schematic representation of the mass-transport mechanism for molecules to be olfactory stimuli. The odorant has to evaporate, enter the nose, reach the olfactory epithelium, adsorb into the olfactory mucosa, enter olfactory receptor binding pockets, and activate an olfactory receptor neuron. Thus, the chemical space of potential odorous compounds is restricted by volatility and polarity constraints.}
    \label{fig:mass_transport}
\end{figure}

Underlying the fragrance perception is the fact that the chemical space of odorous compounds is inherently restricted with mass transport largely determining whether a molecule is odorous or not. \cite{mayhew_2022} To be perceived, molecules must be sufficiently volatile to evaporate, travel the nose, and reach the olfactory epithelium. Yet, they also must possess the right balance of polarity and hydrophobicity to traverse the mucous layer and interact with olfactory receptors to trigger an olfactory receptor neuron. This process is illustrated in \autoref{fig:mass_transport}. 

While simple rules delineate which molecules can be perceived as odorous at all,\cite{mayhew_2022} no comparable general rules with reliable predictive performance exist for more complex qualities of smell, such as odor similarity, intensity, or character, the latter often being described with words like fruity, floral, or rose. Addressing these high dimensional structure-perception relationships requires data‑driven approaches, and has motivated the application of machine learning to predict odor similarity\cite{tom_2025} or character\cite{lee_2023, wang_2022, sharma_2021, sisson_2025} so far. For example, Lee \textit{et al.} \cite{lee_2023} developed a principal odor map based on a message-passing neural network. The model was trained on odor character descriptions and generalized on odor thresholds and odor similarity. Sisson \textit{et al.} \cite{sisson_2025} extended this approach to binary perfumery blends.
As these studies show, the predominant focus in this direction is on the character of the odor rather than its strength even though odor intensity is a decisive factor in the perception of odor. This lack of studies is further reflected in the larger data sets of thousands of compounds available containing descriptive language of odorants, such as data sets like Good Scents \cite{GoodScentsCompany} or Leffingwell \cite{leffingwell&associates_2025}. Only a few studies recorded odor intensity-perception data of molecules with less than 600 different investigated substances in total. \cite{moskowitz_1976, keller_2012, keller_2017,
wakayama_2019, ravia_2020, ma_2021, bierling_2025} 

With respect to the scarcity of intensity data, current state-of-the-art approximations are simple models to predict the psychophysical intensity curve of an individual odorant. Examples are linear (odor values) \cite{fechner_1860}, exponential (Stevens' law), \cite{stevens_1957} or parabolic (e.g. Hill's model) \cite{chastrette_1998b} approaches. However, such models face several limitations, such as limited accuracy for predictions at high concentrations of odorants due to the missing modeling of receptor saturation (linear and exponential) and being based on highly variable odor detection thresholds (linear and exponential). \cite{gemert_2011} The latter does not necessarily equal perceived intensity. \cite{audouin_2001} Recent work has begun to address this issues by predicting parabolic psychophysical curve parameters and extending these predictions to mixtures, using a set of 62 distinct molecules.\cite{pellegrino_2025}


To overcome the scarcity of data and allow for machine learning training on odor strength, we introduce the first odor strength data set containing over 2,000 molecules that allows for generalization across a variety of odor components. Therefore, data from two different sources, namely the Good Scents Company \cite{GoodScentsCompany} and PubChem \cite{PubChem}, were curated and combined.
Using these data, we further investigated the capacity of different descriptors and regressors to predict the odor strength. To our knowledge, these are the first machine learning-based models to predict odor strength categories as an estimate of the odor intensity from molecular structures. An overview of the process is illustrated in \autoref{fig:technical_figure}. 

\section{Results and Discussion}

\subsection{Curated data set and odorous chemical space}

\begin{figure}[!h]
    \centering
    \includegraphics[scale=1]{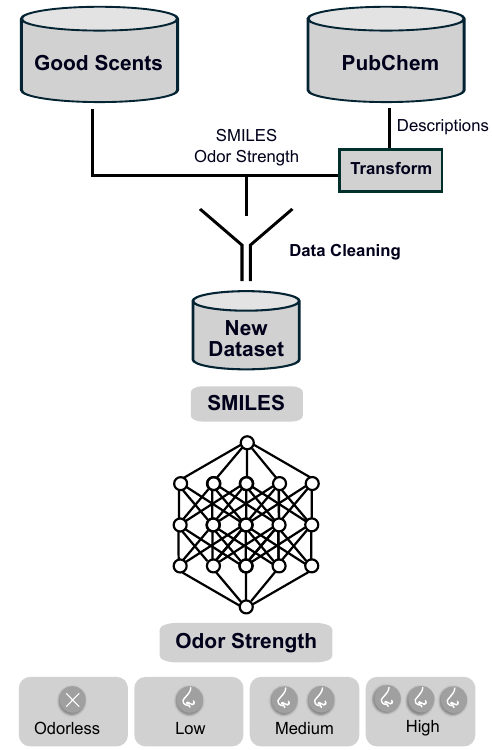}
    \caption{Overview of the developed method. A data set of the ordinal odor strength of more than 2000 compounds was compiled from Good Scents and PubChem. Ordinal regression with a range of state-of-the-art machine learning algorithms was performed to categorize substances by their odor strength.}
    \label{fig:technical_figure}
\end{figure}

Most odor data is corporate property of fragrance houses and odor intensity data is only compiled in data sets for some hundreds of compounds. \cite{moskowitz_1976, keller_2012, keller_2017, wakayama_2019, ravia_2020, ma_2021, bierling_2025} However, machine learning algorithms usually require large data sets for training. Therefore, the first step in our study was to curate a data set. This step is represented in \autoref{fig:technical_figure} at the top. In particular, we evaluated the propensity to combine different sources to compile a larger data set of odor strengths than currently available. As can be seen, we cleaned and combined data from The Good Scents Company \cite{GoodScentsCompany}
and PubChem \cite{PubChem}. 
Data cleaning and preprocessing has to be performed as data from PubChem was labeled by intensity descriptions that had to be transformed to odor strengths. In addition, prior to merging the data, data points were removed that did not represent a valid SMILES string. Details on this process are provided in the Computational Details section.

\begin{figure*}[!h]
    \centering
    \includegraphics[scale=1]{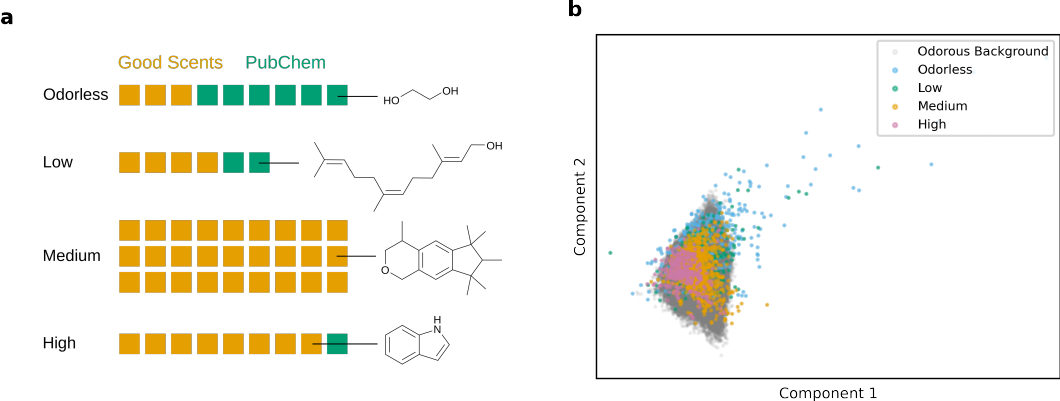}
    \caption{Data set representations. (a) Amount of data for each odor strength. Each square corresponds to about 40 data instances. For each odor strength category, an example molecule is shown. The table with values is shown in the SI in Table S1. (b) 2D PCA of the RDKit descriptors of our curated data set colored by their odor strength and the odorous background data set consisting of 52457 molecules (grey) obtained from a downsample of the GDB-17 database \cite{ruddigkeit_2012} with a predicted odor probability of 50\% or more according to the best-performing model from Mayhew \textit{et al.} \cite{mayhew_2022}.}
    \label{fig:data set}
\end{figure*}

The total data set contains 2,056 molecules. The odor strength distribution for each category, including representative example molecules, is illustrated in \autoref{fig:data set}a. Each block in the figure corresponds to 40 data instances. The exact counts are provided in the Supporting Information (SI) in Table S1. Orange blocks represent data from Good Scents and green blocks indicate data from PubChem. The combination of data should balance the different categories for learning as Good Scents contains mainly medium and high odor strength molecules while PubChem contains mainly low or odorless strength components. 
Despite combining sources to improve balance, medium intensity remains the majority class and low intensity accounts for 12\% of the total, which reflects the underlying availability of annotated strength labels rather than curation bias. A higher balance in odor strength is expected to increase the robustness of the trained model's performance. Additional descriptor repositories (such as those from Leffingwell \cite{leffingwell&associates_2025} or Thiboud \cite{thiboud_1994a}) were not merged because they provide odor character and performance notes but little to no odor strength descriptions, making harmonization across ordinal categories infeasible without unverifiable assumptions.
Similarly, psychophysical intensity data sets were excluded because their ratings are explicit functions of concentration, solvent, and panel protocol; mixing them without a shared concentration scale or covariate model would confound structure–perception relationships and inject systematic bias into ordinal labels.\cite{moskowitz_1976, keller_2012, keller_2017, wakayama_2019, ravia_2020, ma_2021, bierling_2025} This conservative choice defines a single-label, concentration-agnostic ordinal task anchored in molecular structure, while deferring integration of concentration- and solvent-explicit studies to future work where dilution and solvents can be modeled as covariates informed by established psychophysical laws of odor intensity.

To characterize how curated molecules populate an odorous chemical space, the data set was embedded with principal component analysis (PCA) \cite{pearson_1901, hotelling_1933} on RDKit descriptors,\cite{RdkitRdkit2025_03_5} with projections shown in \autoref{fig:data set}b. RDKit descriptors, in this case, comprised 217 structural, physicochemical, and topological parameters of molecules, such as molecular weight, octanol-water partition coefficient (LogP), or the number of heteroatoms. PCA maps this correlated descriptor space to orthogonal principal components ranked by explained variance, enabling faithful visualization on a reduced set of axes for inspection.
For context, an odorous background of 52,457 compounds was constructed by downsampling GDB‑17 and retaining molecules with predicted odorous probability larger than 50\% according to the best-performing model of Mayhew \textit{et al.}\cite{mayhew_2022} Comparable qualitative structure is recovered when (i) excluding the odorous background, (ii) substituting UMAP (Uniform Manifold Approximation and Projection for Dimension Reduction),\cite{mcinnes_2018} a non-linear dimensionality reduction method, for PCA, and (iii) swapping RDKit descriptors for circular fingerprints (Morgan bit- and count-based), with results of all alternatives reported in the SI in Figures S1-S3.
As can be seen from \autoref{fig:data set}b some odorless and low-strength entries fall outside the background space, consistent with mass-transport constraints that bound olfactory space.
Coverage is broad but not uniform; PubChem-derived entries concentrate in specific areas, reflecting the data set’s deliberate emphasis on perfumery-relevant chemotypes. The PCA and UMAP representations visualized by data source are shown in the SI in Figure S3.


Another salient pattern is that odor strength categories do not resolve into distinct clusters but instead overlap substantially in the descriptor space. This makes conventional clustering algorithms, that do not use labels but cluster data solely on input features, inaccurate for separating molecules by their odor strength. We justified this assumption by evaluating several clustering algorithms, including K-means,\cite{lloyd_1982} Gaussian mixture models,\cite{hand_1989} density-based spatial clustering of applications with noise (DBSCAN), \cite{ester_1996} spectral,\cite{jianboshi_2000, NIPS2001_801272ee} and agglomerative\cite{ward_1963} clustering. No groups corresponding to odor strength were formed. Additional information can be found in the SI in Section S1.2, including evaluation metrics in Table S2 and the best clustering result in Figure S6. 

To analyze which features are most important for data separation, the feature importance of the principal components was analyzed. The corresponding PCA loadings indicate that the first principal component is primarily influenced by features, which describe molecular weight, size, and connectivity. The second principal component reflects contributions from heteroatoms and polarity. The exact distribution of the top 15 features to the different principal components can be found in the SI in Tables S3 and S4. These features are major factors governing mass transport of molecules from odor source to olfactory receptors. 
This observation aligns with findings of Mayhew \textit{et al.}, who showed that the mass transport is key for determining whether molecules are odorous or not.
Consequently, the chemical space occupied by odorous molecules becomes progressively narrower as odor strength increases, which reflects the tighter constraints imposed by volatility and polarity. Further visualization plots underlying this claim can be found in the SI in Figures S4 and S5.

\subsection{Odor strength predictors and model validation}

As conventional cluster algorithms fail to group the data by their odor strength, further supported by dimensionality reduction and chemical space analysis in the previous section, supervised learning algorithms are used for odor strength prediction. In contrast to unsupervised learning models (clustering), supervised learning models utilize labeled data to guide the learning process. Consequently, the objective was to determine whether such a model could learn the relationship between molecular structure and odor strength of the curated data.

\begin{figure*}[!h]
    \centering
    \includegraphics[scale=1]{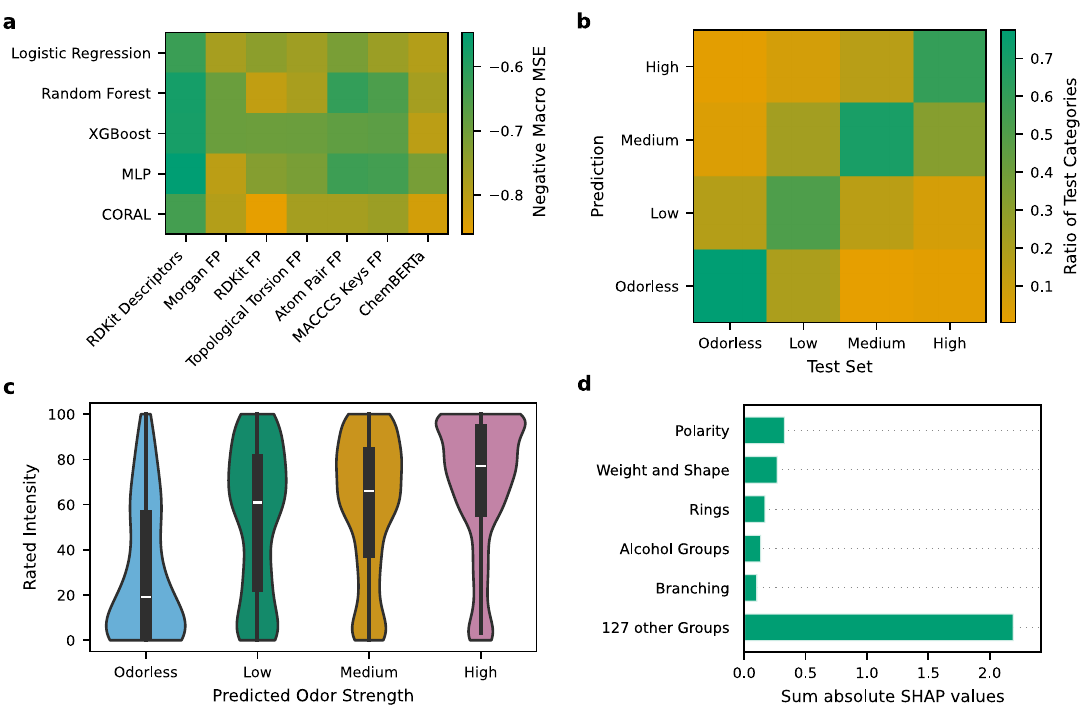}
    \caption{Best-performing model performance for the direct prediction approach. (a) Macro averaged mean squared error (MSE) across odor strength categories in  the validation sets obtained from cross-validation of the best models for all combinations of molecule encoders (bottom) and predictors (left). MLP is multi-layer-perceptron and FP fingerprint. 
    (b) Confusion matrix normed by the number of test samples for the best-performing model on the test set, averaged over 10 random-seeded training runs.
    (c) Area-normed violin plots of the best-performing model predictions for novel molecules from Keller \textit{et al.}\cite{keller_2017} compared with their experimentally rated odor intensities (from 0 to 100; 13-108 per molecule) at 1/1000 dilution. (d) Global SHAP (SHapley Additive exPlanations)\cite{lundberg2017unified} feature importance of the most influential feature groups of the best-performing model. The RDKit descriptor features were grouped using agglomerative clustering based on their feature value correlation (threshold: 0.75). The absolute SHAP values within each group were summed.}
    \label{fig:model_performance}
\end{figure*}

To predict odor strength based on molecular representations, two distinct modeling strategies were tested: First, we used a "direct approach". For this, we defined four classes, \textit{i.e.}, odorless, low, medium, and high odor strength, followed by training a model on all of these classes. Second, we separated odorless molecules from odorous substances. This approach then requires two predictions, one that decides whether a molecule smells or not and the second should categorize the smell into the strength of odor (low, medium, high). Both strategies were evaluated to test complementary hypotheses about the structure-to-perception pathway: a single-task ordinal learner might best capture global trade-offs across all classes, while a hierarchical, two-step pipeline could exploit different molecular features for mass transport first and for receptor interaction second.

Five regression algorithms were tested, in particular classical logistic regression,\cite{verhulst_1845} random forest,\cite{breiman_2001} extreme gradient boosting (XGBoost),\cite{chen_2016} multi-layer-perceptrons (MLP),\cite{rumelhart_1986, adam_2019} and consistent rank logits (CORAL),\cite{cao_2020} an MLP architecture designed for ordinal regression. Each algorithm was tested with seven widely used molecular encoding strategies to represent a molecule. This breadth allowed us to assess the predictive performance in fundamentally different feature spaces. We calculated structural, topological, and physicochemical descriptors (RDKit descriptors\cite{RdkitRdkit2025_03_5}) and employed classical fingerprints that highlight subtle structural differences. These included the circular substructure Morgan fingerprint,\cite{morgan_1965, rogers_2010} the predefined substructure MACCS keys fingerprint,\cite{durant_2002} and several topological fingerprints that capture sequences of atom connectivity for longer-range relationships, such as the RDKit fingerprint,\cite{RdkitRdkit2025_03_5} the topological torsion fingerprint,\cite{nilakantan_1987} and the atom pair fingerprint.\cite{carhart_1985} Beyond these classical approaches, we evaluated more recent representation learning techniques. ChemBERTa-2,\cite{ahmad_2022} a language model pretrained on 77 million SMILES strings provided data-driven embeddings. Additionally, graph-based encodings using message passing neural networks have recently demonstrated promising results to predict odor character \cite{lee_2023} and similarity.\cite{tom_2025} Consequently, we applied ChemProp,\cite{yang_2019, heid_2024} a framework for message passing neural networks. Recognizing the potential of pretraining to potentially improve downstream model performance, \cite{ramanujan_2023}  we further tested CheMeleon,\cite{burns_2025} a foundational ChemProp model pretrained on Mordred descriptors of 77 million molecules.
All model hyperparameters were optimized using 10 times repeated 10-fold cross validation. The high number of repetitions reduced the impact of noise in the data. The hyperparameters that were optimized are provided in the SI in section S2 (Tables S5-S14) for the different models used.
We chose the macro averaged mean squared error (macro MSE) as the average of the MSEs computed for each odor strength categories. The equation is provided in the Computational Details section in \autoref{eq:macro_mse}. The macro MSE penalizes larger errors more heavily while equally weighting all odor strength categories, regardless of how many samples they contain. In contrast, the common micro MSE is calculated globally across all classes and dominated by the majority classes.

The best-performing models correspond to the direct prediction approach and results are shown in \autoref{fig:model_performance}a. The results of the indirect approach steps are provided in the SI in Figures S7-S12 and are generally, but not significantly, less accurate. The difference in the macro MSE values for each model and descriptor combined is plotted in Figure S10 with a comparison of the best predictors for each approach for each odor strength category in Figure S12.
Specifically, the combined best-performing model of the indirect approach (XGBoost with RDKit descriptors and XGBoost with topological torsion fingerprint) performs only slightly worse (0.06 macro MSE difference; macro MSE=0.59 and the R\textsuperscript{2}=0.53) than the best-performing model (MLP with 217 RDKit descriptors) of the direct approach (macro MSE=0.53$\pm$0.03, R\textsuperscript{2}=0.57$\pm$0.02) on the same hold-out test set for an average of 10 trained models.
The normed confusion matrix of the best model of the direct approach is shown in \autoref{fig:model_performance}b (indirect approach in the SI in Figure S11). Notably, the highest accuracy was observed for the medium odor strength class, which was most prevalent in both the training and test set. Most misclassifications occurred between adjacent odor strength categories. Overall, the best-performing model correctly classified more than 60\% of the test instances within their respective categories. The indirect model demonstrated slightly better classification of odorless, medium and high odor strengths, but its predictions for low odor strengths were both less frequent and substantially less accurate. 
Given the reduced potential for error propagation, lower computational resource consumption during both training and prediction and generally higher performance across models and descriptors, the remainder of this study will focus on the direct approach.

Finally, we assessed the model performance on another literature-derived test set that contains experimental odor intensity ratings.\cite{keller_2017} The molecules were entirely distinct from the data set derived and used above, with no close structural similarities evaluated by the Tanimoto similarity of the bit-based Morgan fingerprints. \autoref{fig:model_performance}c displays the results using a violin plot that shows the mean of the rated intensity and its distribution among the intensity ratings (white line and black box) in addition to the distribution (shape, color) of the predictions. As can be seen, predicted odor strengths correlate well with the rated odor intensities (ranging from 0 to 100).
Despite considerable variability among individual ratings our model provides a reasonable approximation of perceived odor strength. This trend is further confirmed for different dilutions of the same substances, \textit{i.e.}, a 1/10\textsuperscript{5} dilution with results shown in the SI in Figure S13.

\subsection{Physicochemical determinants of odor strength}

While the performance of a machine learning model is a central aspect that can be assessed, for instance via MSEs or other metrics, model explainability is equally important.\cite{coussement_2024} To provide insights into the factors that drive model decisions, we use SHAP (SHapley Additive exPlanations),\cite{lundberg2017unified} which assigns each feature a contribution value to a model's prediction based on Shapley values \cite{shapley_1953} from cooperative game theory. This approach provides both global and local interpretability of the model's behavior. To assess global feature importance, we computed the absolute SHAP values across all features and aggregated them into groups based on agglomerative clustering\cite{ward_1963} on their feature value correlation (threshold of 0.75). The resulting SHAP-based feature importance for the most influential feature groups is shown in \autoref{fig:model_performance}d and the contributions of each feature to its corresponding group in the SI in Figures S15-S19.

In agreement with the loadings of the first two principal components of the described PCA (\autoref{fig:data set}b), features related to molecular polarity, such as the number of hydrogen acceptors or donors and the number of heteroatoms, as well as features describing molecular weight and shape, such as molecular weight or Chi descriptors,\cite{hall_1991} exhibit the highest importance in the model's predictions. Additionally, ring-related properties, the presence of alcohol groups and branching patterns contributed substantially to the model predictions. Notably, the cumulative impact of the remaining feature groups was markedly higher, reflecting the complexity of the model's decision process.
Furthermore, the feature importance per odor strength is shown in the SI in Figure S20
No clear trends that specific properties are more relevant for higher or lower odor strengths could be observed.
Representative examples of local feature group contributions for four molecules representing each of the odor strength classes are shown in the SI in Figure S21. These examples are in line with results found globally and illustrate that polarity, molecular weight and shape substantially influence odor strength. 
 
\section*{Conclusion}

In this work, the propensity of machine learning for the prediction of odor strength was assessed. We show that odor strength can be predicted directly from molecular structure using a newly curated data set of more than 2,000 odorants compiled and merged from two different sources. 
Two-dimensional reduced representations of the chemical odorous space showed that odor strength categories overlap significantly and do not form specific clusters when using molecular descriptors or circular fingerprints as representations, hence making unsupervised learning methods difficult. Key variation drivers to distinguish classes in odor strength were analyzed with mass-transport-related features such as molecular weight, size, shape and polarity, being more restrictive for higher odor strengths.
Comprehensive benchmarking of state-of-the-art molecular encodings and predictive algorithms identified a mulit-layer perceptron (MLP) trained directly on RDKit descriptors as the best-performing model. We tested two different learning strategies: first, we tested the approach of directly predicting odorless, low, medium, and high odor strength categories. Second, we tested the separation of odorless molecules and odorants first, classifying odorants by their odor strength in a second step. While both models exhibit comparable results, the first, direct approach is more stable and generally more accurate.
Our proposed model can be utilized to predict the odor strength of novel molecules and thus supports rational fragrance design, where weighting of components by odor strength could be beneficial.

One of the main challenges in the prediction of odor strengths remains the labeling of molecules, which is highly subjective and requires evaluations of many individuals for robust results. However, odor perception is highly variable between individuals. \cite{keller_2012}
Moreover, the discrete classification itself neglects continuous variations in odor intensity within an odor strength category.
Potentially one of the most critical concerns in most odor literature data is the purity of fragrance chemicals. Impurities can alter the odor even in very low concentrations. \cite{paoli_2017} A recent study using gas chromatography-olfactometry reported that 22\% of the supposedly odorous molecules investigated were, in fact, odorless. \cite{mayhew_2022}
To further advance odor intensity modeling, there is a need for comprehensive odor intensity data covering a wider range of molecules and mixtures, measured at multiple concentrations and with well-characterized impurities. Such data would support models that better capture the empirically known shape of the monotonic, sigmoidal relationship between the logarithm of odorant concentration and perceived odor intensity. whose slopes vary for different odorants. \cite{mainland_2014} While odorousness\cite{mayhew_2022} and odor strength can be reliably predicted from structure-related transport properties, incorporating biological information, such as receptor responses, may yield even more accurate and mechanistically grounded models. 
As such, this work provides a step towards data-driven in silico fragrance design.

\section*{Author contributions}
PF: Data curation, machine learning training and validation, analysis, manuscript writing and revision.\\
JW: Conceptualization, analysis, manuscript writing and revision.

\section*{Conflicts of interest}
There are no conflicts to declare.

\section*{Data availability}


This study was carried out using publicly available data from the Good Scents website \cite{GoodScentsCompany} and PubChem \cite{PubChem}.
The raw Good Scents data set is protected under the U.S. and Foreign Copyright Laws. Consequently, the data set is not provided; however the code to obtain it is available.
The raw PubChem data set and the code for webscrapping the raw data sets, creating and cleaning the curated data set, analyzing the chemical space, training and validating the models, revealing feature importance and applying the best-performing model with an interactive notebook to make new predictions can be found at \url{https://github.com/peter-fichtelmann/odor-strength} and is uploaded at Zenodo under DOI:10.5281/zenodo.17828661 (version 1.0.4).

\section*{Acknowledgements}

The authors gratefully acknowledge funding from Bell Flavors \& Fragrances GmbH, Leipzig.
Computations for this work were done using resources of the Leipzig University Computing Center.

\section*{Computational Details}

\subsection*{Data set generation and analysis}

For data curation, a web crawler was built using the python package BeautifulSoup\cite{richardson_2025} to extract the odor strength, SMILES strings, and CAS-numbers of all 42,234 entries on The Good Scents Company website.\cite{GoodScentsCompany} Due to some incorrect SMILES in the Good Scents repository,  CAS-numbers were first converted into SMILES using the CAS Common Chemistry API \cite{americanchemicalsociety_2025}. If this conversion turned out unsuccessful, SMILES strings from Good Scents were used. If still unavailable, CAS-numbers were converted to SMILES via PubChem.\cite{PubChem} Since only 13 compounds were annotated with 'very high' odor strength, they were reclassified as 'high'. The specific compounds are highlighted in section S6 in Table S15.
Regarding the PubChem data, all compound IDs (CIDs) of entries with odor descriptions (2,357) according to the PubChem classification browser were retrieved, and corresponding SMILES and odor descriptions were obtained via web crawling. 
All SMILES strings were canonicalized using RDKit.\cite{RdkitRdkit2025_03_5}
Odor descriptions from PubChem were converted into odor strength categories if predefined keywords were present in the description. The keywords for each category are provided in the SI in section S6. Only molecules with valid canonicalized SMILES were retained. Duplicated SMILES entries from PubChem were removed. A total of 332 SMILES containing dots were identified. Dots in SMILES are not part of a molecule’s covalent structure, but indicate separate disconnected fragments, e.g. ions or isomers. 15 entries representing isomers with SMILES separated by dots were split into individual instances of the same odor strength. The remaining SMILES with dots and the ambiguous entries of 'carob bean absolute' and 'galbanum resinoid' were excluded. Furthermore, entries with SMILEs including other elements than H, C, O, N, S, Cl were discarded. To bias the data set towards perfumery materials, PubChem compounds with a maximum Tanimoto similarity below 20\% (based on their bit-based Morgan fingerprints, radius=3, nBits=2048) to any Good Scents compound were removed. In this way, a total of 2056 data points were collected with 1684 entries from Good Scents and 372 entries from PubChem.

In addition, another independent hold-out test set was generated using data from Keller \textit{et al.},\cite{keller_2017} downloaded via Pyrfume.\cite{hamel_2024} Molecules with 80\% or more respective Tanimoto Similarity of their bit-based Morgan fingerprints (radius=3, nBits=2048) to at least one molecule in the train set were removed to ensure that the test set contained only novel compounds with proper dissimilarity to trained compounds. Several dilutions of compounds were available. Only those with more than 35 molecules per dilution were considered. The 1/1000 and 1/10\textsuperscript{5} dilutions comprised 126 and 123 unique molecules, respectively. Each molecule was rated between 13 and 108 times (average 48 (1/1000), 39 (1/10\textsuperscript{5})) by independent laymen panelists providing a mean and standard deviation of rated intensities for each molecule per dilution.


\subsection*{Model training and validation}
For training machine learning models, the whole data set was split into 80\% training set and 20\% hold-out test set as implemented in scikit-learn.\cite{pedregosa_2011} The training set was used for 10-fold cross-validation over 10 random seed repetitions. The training set was split for each run into 90\% training and 10\% validation set to find optimal hyperparameters. The high number of repetitions was chosen to reduce chance-based variation. All splits were stratified to retain approximately the same percentage of odor strength categories in each subset. In addition, molecules with 80\% or more respective Tanimoto similarity of their bit-based Morgan fingerprints (radius=3, nBits=2048) were grouped in the same split subsets to avoid leakage of structurally very similar compounds across sets.

The macro MSE computed across each odor strength category was used as evaluation metric, following the equation:

\begin{equation}
    macro\text{\:}MSE = \frac{1}{N} \sum^N_{i=1}MSE_i
    \label{eq:macro_mse}
\end{equation}

where $N$ is the number of categories (four) and $MSE_i$ the MSE for category $i$.

\subsubsection*{Hyperparameter optimization}

Hyperparameters of each model were tuned over 100 trials with Optuna \cite{akiba_2019} using the tree-structured Parzen estimator. In each trial, the previously described cross-validation procedure was applied. Detailed hyperparameter ranges and sampling strategies for each molecular encoder and predictor are provided in the SI in Section S2 in Tables S5-S14.
Two distinct objective metrics were employed: 1) for the direct and the second step of the indirect approach: the macro-averaged MSE across odor strength categories of the validation set; 2) for the first step of the indirect approach (binary classifier: if a molecule is odorous): F1-score where the target was the minority class.
To address class imbalance, all predictors used cost-sensitive learning via odor strength category weighted loss functions.
A 25-percentile pruner discarded unpromising trials with a deviation larger than a tolerance interval (macro MSE= 0.02, F1 score=0.015) to the best trial after each cross-validation repetition.

\subsection*{SHAP feature importance analysis}

SHAP values of the test set were computed using the training set as a background to estimate average and sample feature values using the python SHAP\cite{lundberg2017unified} package. Several RDKit descriptor feature values were highly correlated (correlation matrix provided in the SI in Figure S14).
Consequently, we applied agglomerative clustering\cite{ward_1963} to group highly correlated features (correlation threshold 0.75) into 148 groups. While methods exist to account for feature correlations during SHAP value estimation,\cite{aas_2021, hothorn_2006, olsen_2022} the computational cost of conditional or dependence‑aware SHAP grows exponentially regarding the number of features.\cite{olsen_2024} Regarding our number of features and number of groups, this is currently not feasible. Therefore, we report group‑aggregated importance by summing SHAP values within correlation‑based clusters and interpret effects at the group level.



\balance



\providecommand*{\mcitethebibliography}{\thebibliography}
\csname @ifundefined\endcsname{endmcitethebibliography}
{\let\endmcitethebibliography\endthebibliography}{}

\end{document}


\pagestyle{fancy}
\thispagestyle{plain}
\fancypagestyle{plain}{
\renewcommand{\headrulewidth}{0pt}
}

\makeFNbottom
\makeatletter
\renewcommand\LARGE{\@setfontsize\LARGE{15pt}{17}}
\renewcommand\Large{\@setfontsize\Large{12pt}{14}}
\renewcommand\large{\@setfontsize\large{10pt}{12}}
\renewcommand\footnotesize{\@setfontsize\footnotesize{7pt}{10}}
\makeatother

\renewcommand{\thefootnote}{\fnsymbol{footnote}}
\renewcommand\footnoterule{\vspace*{1pt}%
\color{cream}\hrule width 3.5in height 0.4pt \color{black}\vspace*{5pt}} 
\setcounter{secnumdepth}{5}

\makeatletter 
\renewcommand\@biblabel[1]{#1}            
\renewcommand\@makefntext[1]%
{\noindent\makebox[0pt][r]{\@thefnmark\,}#1}
\makeatother 
\renewcommand{\figurename}{\small{Fig.}~}
\sectionfont{\sffamily\Large}
\sectionfont{\normalsize}
\subsectionfont{\bf}
\setstretch{1.125} 
\setlength{\skip\footins}{0.8cm}
\setlength{\footnotesep}{0.25cm}
\setlength{\jot}{10pt}
\titlespacing*{\section}{0pt}{4pt}{4pt}
\titlespacing*{\section}{0pt}{15pt}{1pt}

\fancyfoot{}
\fancyfoot[RO]{\footnotesize{\sffamily{1--\pageref{LastPage} ~\textbar  \hspace{2pt}\thepage}}}
\fancyfoot[LE]{\footnotesize{\sffamily{\thepage~\textbar\hspace{3.45cm} 1--\pageref{LastPage}}}}
\fancyhead{}
\renewcommand{\headrulewidth}{0pt} 
\renewcommand{\footrulewidth}{0pt}
\setlength{\arrayrulewidth}{1pt}
\setlength{\columnsep}{6.5mm}
\setlength\bibsep{1pt}

\makeatletter 
\newlength{\figrulesep} 
\setlength{\figrulesep}{0.5\textfloatsep} 

\newcommand{\topfigrule}{\vspace*{-1pt}%
\noindent{\color{cream}\rule[-\figrulesep]{\columnwidth}{1.5pt}} }

\newcommand{\botfigrule}{\vspace*{-2pt}%
\noindent{\color{cream}\rule[\figrulesep]{\columnwidth}{1.5pt}} }

\newcommand{\dblfigrule}{\vspace*{-1pt}%
\noindent{\color{cream}\rule[-\figrulesep]{\textwidth}{1.5pt}} }

\makeatother

\renewcommand*\rmdefault{bch}\normalfont\upshape
\rmfamily
\section*{}
\vspace{-1cm}

\renewcommand{\figurename}{Figure}
\renewcommand{\thefigure}{S\arabic{figure}}
\renewcommand{\thetable}{S\arabic{table}}
\renewcommand{\thesection}{S\arabic{section}}
\renewcommand{\thesubsection}{S\arabic{section}.\arabic{subsection}}
\renewcommand{\thesubsubsection}{S\arabic{section}.\arabic{subsection}.\arabic{subsubsection}}

\begin{center}
\LARGE\textbf{Supporting Information for "Machine learning for smell: Ordinal odor strength prediction of molecular perfumery components"}
\end{center}

 \vspace{3mm}

\begin{center}
    \large{Peter Fichtelmann\textit{$^{a}$} and Julia Westermayr\textit{$^{a,b}$}}
\end{center}

\vspace{3mm}

\begin{center}
    $^{a}$~Wilhelm-Ostwald Institute of Physical and Theoretical Chemistry, Leipzig University, Linn\'{e}stra{\ss}e 2, 04103 Leipzig, Germany \\
    $^{b}$ Center for Scalable Data Analytics and Artificial Intelligence Dresden/Leipzig, Humboldtstra{\ss}e 25, 04105 Leipzig, Germany
\end{center}

\vspace{10mm}

\tableofcontents
\clearpage




\section{Curated data set and odorous chemical space details}

The number of molecules in the curated data set regarding their odor strength categories and data set source is specified in \autoref{tab:molecules}.


\begin{table}[h]
    \small
    \centering
    \caption{\ Number of molecules in the data set by odor strength. 12 Good Scents molecules with very high odor strength were counted towards high odor strength.}
    \centering
    \begin{tabular}{lrrr}
    \hline
    Odor strength & Count Good Scents & Count PubChem & Total count \\
    \hline
    odorless & 144 & 239 & 383 \\
    low & 158 & 86 & 244 \\
    medium & 1071 & 0 & 1071 \\
    high & 311 & 47 & 358 \\
    Sum & 1684  & 372 & 2056 \\
    \hline
    \end{tabular}
    \label{tab:molecules}
\end{table}

\clearpage

\subsection{2D representations of the odorous chemical space}

\begin{figure}[h]
    \centering
    \includegraphics[scale=1]{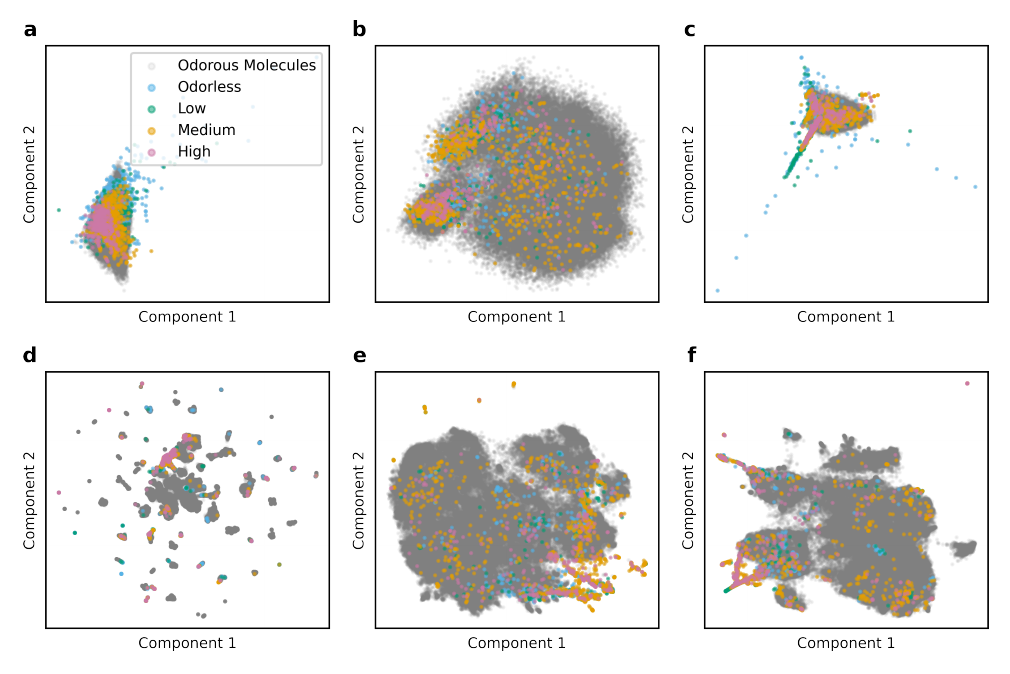}
        \caption{2D representation of our curated data set colored by their odor strength and the 52457 molecules (grey) of a downsample from the GDB-17 database \cite{ruddigkeit_2012} with a predicted odor probability of 50\% or more according to the best-performing model from Mayhew \textit{et al.} \cite{mayhew_2022} using (a) PCA on RDKit Descriptors, (b) PCA on the bit-based Morgan fingerprint, (c) PCA on the count-based Morgan fingerprint, (d) UMAP on RDKit Descriptors, (e) UMAP on the bit-based Morgan fingerprint, (f) UMAP on the count-based Morgan Fingerprint. All Morgan fingerprints used a radius of 3 and a length of 2048. UMAP was performed with default hyperparameters.}
    \label{fig:pca_umap_odor_strength_w_background}
\end{figure}

\begin{figure}[h]
    \centering
    \includegraphics[scale=1]{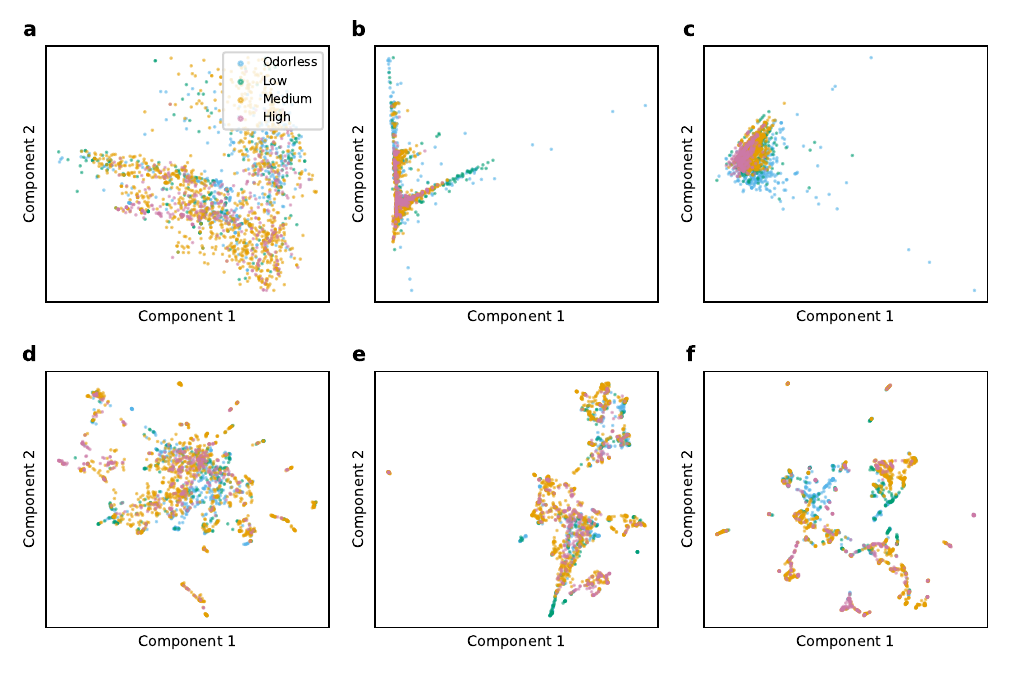}
        \caption{2D representation of our curated data set colored by their odor strength without an additional odor molecule background using (a) PCA on RDKit Descriptors, (b) PCA on the bit-based Morgan fingerprint, (c) PCA on the count-based Morgan fingerprint, (d) UMAP on RDKit Descriptors, (e) UMAP on the bit-based Morgan fingerprint, (f) UMAP on the count-based Morgan Fingerprint. All Morgan fingerprints used a radius of 3 and a length of 2048. UMAP was performed with default hyperparameters.}
    \label{fig:pca_umap_odor_strength}
\end{figure}

\begin{figure}[h]
    \centering
    \includegraphics[scale=1]{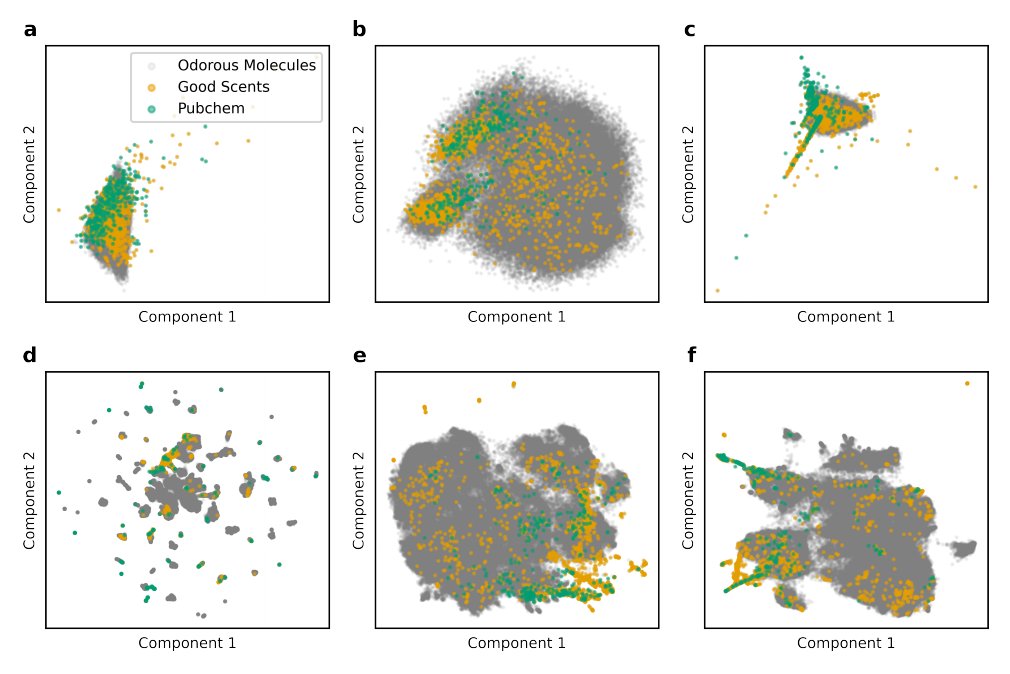}
        \caption{2D representation of our curated data set colored by their source data set and the 52457 molecules (grey) of a downsample from the GDB-17 database \cite{ruddigkeit_2012} with a predicted odor probability of 50\% or more according to the best-performing model from Mayhew \textit{et al.} \cite{mayhew_2022} using (a) PCA on RDKit Descriptors, (b) PCA on the bit-based Morgan fingerprint, (c) PCA on the count-based Morgan fingerprint, (d) UMAP on RDKit Descriptors, (e) UMAP on the bit-based Morgan fingerprint, (f) UMAP on the count-based Morgan Fingerprint. All Morgan fingerprints used a radius of 3 and a length of 2048. UMAP was performed with default hyperparameters.}
    \label{fig:pca_umap_gs_pubchem_w_background}
\end{figure}

\begin{figure}[h]
    \centering
    \includegraphics[scale=1]{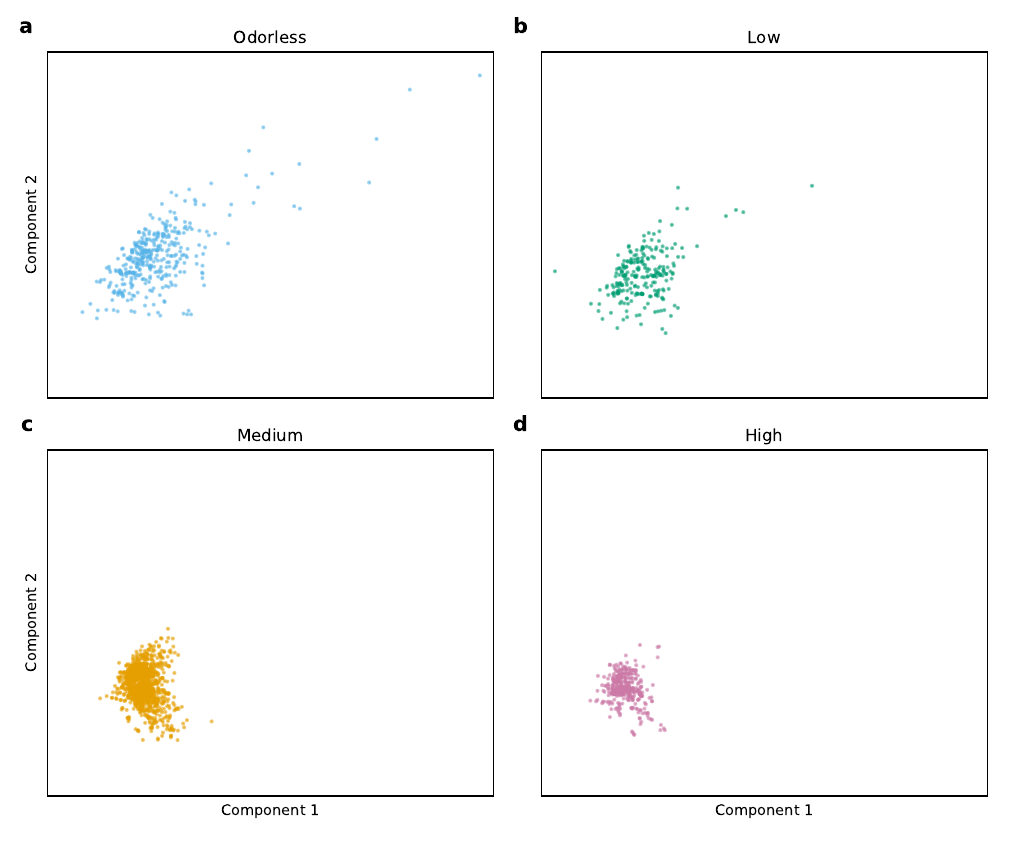}
        \caption{2D PCA of the RDKit descriptors of our curated data set and the 52457 molecules of a downsample from the GDB-17 database \cite{ruddigkeit_2012} with a predicted odor probability of 50\% or more according to the best-performing model from Mayhew \textit{et al.} \cite{mayhew_2022}. (a) Odorless, (b) Low odor strength, (c) Medium odor strength, (d) High odor strength.}
    \label{fig:pca_rdkit_desc_per_odor_strength_w_background}
\end{figure}

\begin{figure}[h]
    \centering
    \includegraphics[scale=1]{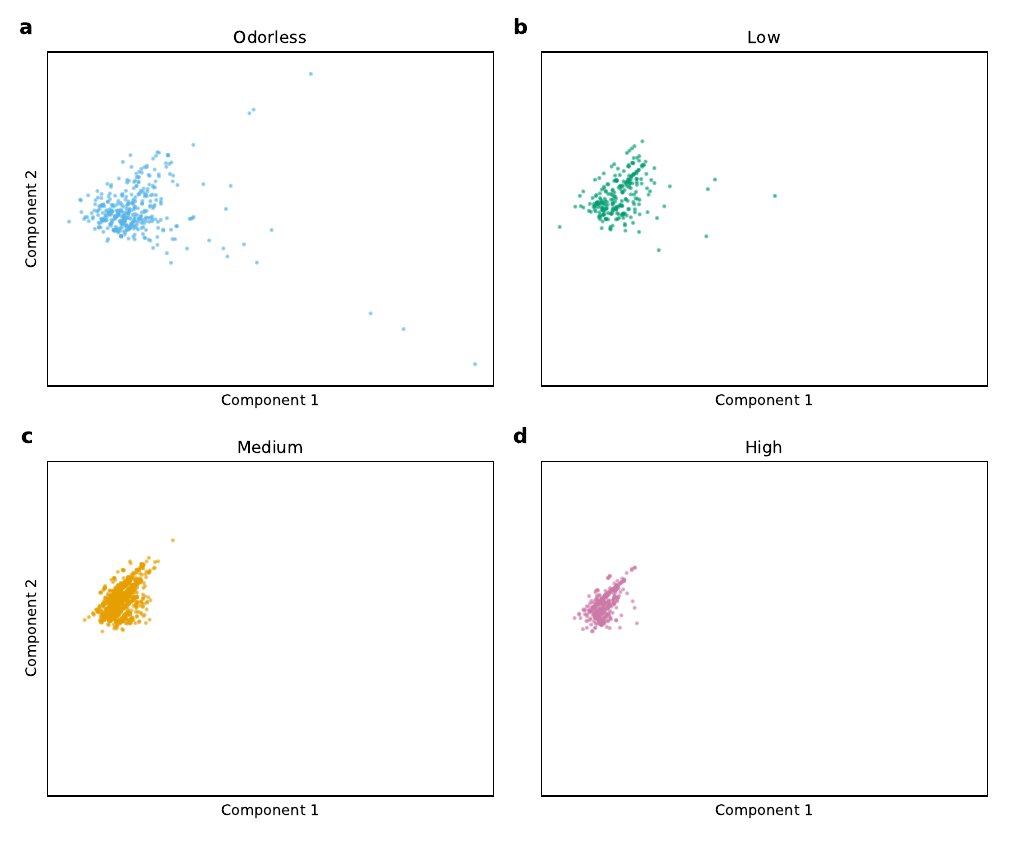}
        \caption{2D PCA of the RDKit descriptors of our curated data set. (a) Odorless, (b) Low odor strength, (c) Medium odor strength, (d) High odor strength.}
    \label{fig:pca_rdkit_desc_per_odor_strength}
\end{figure}

2D representations of the odorous chemical space, including molecules of our data set, are shown in \autoref{fig:pca_umap_odor_strength_w_background}-\autoref{fig:pca_rdkit_desc_per_odor_strength}.
We observed similar results with (\autoref{fig:pca_umap_odor_strength_w_background}, \autoref{fig:pca_umap_gs_pubchem_w_background}, \autoref{fig:pca_rdkit_desc_per_odor_strength_w_background}) and without (\autoref{fig:pca_umap_odor_strength}, \autoref{fig:pca_rdkit_desc_per_odor_strength}) a background of potential odorous molecules (downsample from the GDB-17 database \cite{ruddigkeit_2012} with a predicted odor probability of 50\% or more according to the best-performing model from Mayhew \textit{et al.} \cite{mayhew_2022}).
Principal component analysis (PCA) and uniform manifold approximation and projection (UMAP) were used to reduce the dimensionality of RDKit descriptors and a bit- and count-based Morgan fingerprint(radius=3, nBits=2048) colored by odor strength in \autoref{fig:pca_umap_odor_strength_w_background}, \autoref{fig:pca_umap_odor_strength}, \autoref{fig:pca_rdkit_desc_per_odor_strength_w_background}, \autoref{fig:pca_rdkit_desc_per_odor_strength} and colored by data set source in \autoref{fig:pca_umap_gs_pubchem_w_background}.

\clearpage

\subsection{Clustering}

\begin{figure}[h]
    \centering
    \includegraphics[scale=1]{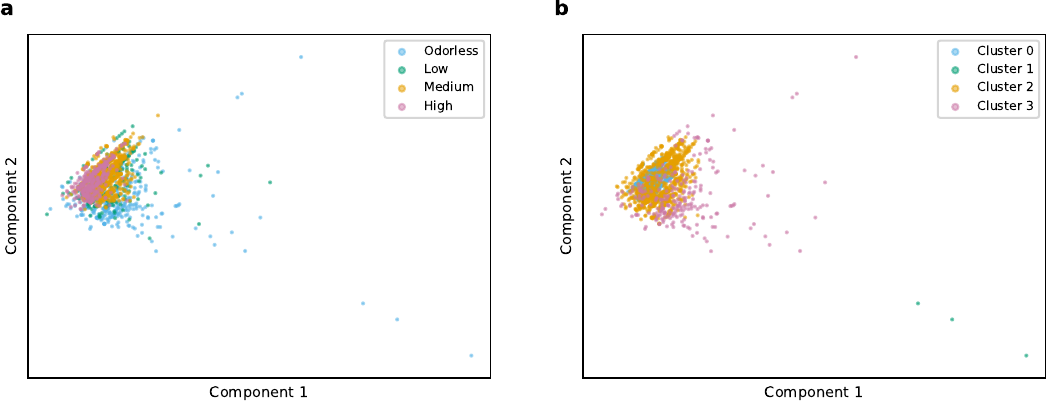}
        \caption{Best clustering result. (a) 2D PCA of the RDKit Descriptors of our curated data set colored by their odor strength. (b) 2D PCA of the RDKit Descriptors of our curated data set colored by the clusters of a gaussian mixture model, our best-performing clustering algorithm (ARI: 0.29).}
    \label{fig:clustering}
\end{figure}

In \autoref{tab:clustering} we report the adjusted rand indices (ARIs)\cite{hubert_1985} and silhouette scores\cite{rousseeuw_1987} on our data set in comparison to the ground truth odor strengths of the investigated clustering algorithms K-means,\cite{lloyd_1982} gaussian mixture models,\cite{hand_1989} density-based spatial clustering of applications with noise (DBSCAN),\cite{ester_1996} spectral\cite{jianboshi_2000, NIPS2001_801272ee} and agglomerative \cite{ward_1963} clustering. For each method, we optimized key hyperparameters with respect to the adjusted rand index (ARI).
The ARI is a chance-corrected measure of clustering performance compared to the ground truth ranging from 0 (random assignment) to 1 (perfect agreement). The silhouette score corresponds to the tightness and separation of the clusters, ranging from -1 to 1 (best 1, 0 overlapping clusters). The silhouette scores reached values up to 1.0, but the ARIs were low with a maximum of 0.29. A 2D PCA RDKit Descriptors representation of the best clustering result regarding the ARI (ARI 0.29, silhouette score: 0.06) is shown in \autoref{fig:clustering}.

\begin{table}[h]
    \small
    \centering
    \caption{\ Adjusted rand index (ARI) and silhouette scores of 5 hyperparameter tuned clustering methods using bit-based (Morgan Binary) and count-based (Morgan Count) Morgan fingerprints (radius=3, nBits=2048) and RDKit Descriptors. The ARI was benchmarked against the odor strengths.}
    \centering
    \begin{tabular}{lrrr}
    \hline
    Descriptor & Algorithm & ARI & Silhouette \\
    \hline
    Morgan Binary & K-Means & 0.021 & 0.012 \\
    Morgan Binary & GMM & 0.024 & 0.011 \\
    Morgan Binary & DBSCAN & 0.011 & 1.000 \\
    Morgan Binary & Spectral & 0.002 & 0.304 \\
    Morgan Binary & Agglomerative & 0.035 & -0.113 \\
    Morgan Count & K-Means & 0.033 & 0.137 \\
    Morgan Count & GMM & 0.071 & 0.146 \\
    Morgan Count & DBSCAN & 0.043 & 0.205 \\
    Morgan Count & Spectral & 0.000 & 0.603 \\
    Morgan Count & Agglomerative & 0.058 & 0.151 \\
    RDKit Descriptors & K-Means & 0.072 & 0.081 \\
    RDKit Descriptors & GMM & 0.285 & 0.122 \\
    RDKit Descriptors & DBSCAN & -0.000 & 0.626 \\
    RDKit Descriptors & Spectral & 0.001 & 0.506 \\
    RDKit Descriptors & Agglomerative & 0.029 & 0.047 \\
    \hline
    \end{tabular}
    \label{tab:clustering}
\end{table}

\clearpage

\subsection{PCA loadings}

The top 15 PCA loadings of the first two principal components are provided without odorous background in \autoref{tab:pca_loadings} and with in \autoref{tab:pca_loadings_w_background}. In both cases, the first principal component can be attributed to molecular weight and shape and the second to heteroatoms and polarity.

\begin{table}[h]
    \small
    \centering
    \caption{\ Top 15 principal components loadings (PC1, PC2) of the principal component analysis on the RDKit Descriptors of our data set.}
    \centering
    \begin{tabular}{lrrr}
    \hline
    Feature PC1 & PC1 Loading & Feature PC2 & PC2 Loading \\
    \hline
    HeavyAtomMolWt & 0.162104 & MolLogP & 0.184090 \\
    ExactMolWt & 0.162090 & VSA\_EState7 & 0.179914 \\
    MolWt & 0.162061 & SlogP\_VSA5 & 0.173863 \\
    NumValenceElectrons & 0.161716 & TPSA & -0.152584 \\
    Chi0 & 0.160954 & VSA\_EState8 & 0.152035 \\
    HeavyAtomCount & 0.160812 & NumHeteroatoms & -0.150515 \\
    Chi1 & 0.159409 & NOCount & -0.145843 \\
    LabuteASA & 0.158483 & SMR\_VSA5 & 0.137586 \\
    Chi0v & 0.153552 & NHOHCount & -0.133345 \\
    Chi0n & 0.153145 & PEOE\_VSA7 & 0.133058 \\
    Kappa1 & 0.151412 & EState\_VSA5 & 0.131315 \\
    Chi1v & 0.151135 & NumHDonors & -0.130139 \\
    Chi1n & 0.150262 & PEOE\_VSA6 & 0.130132 \\
    MolMR & 0.147677 & EState\_VSA8 & 0.129424 \\
    Chi2v & 0.146024 & NumHAcceptors & -0.129146 \\
    \hline
    \end{tabular}
    \label{tab:pca_loadings}
\end{table}

\begin{table}[h]
    \small
    \centering
    \caption{\ Top 15 principal components loadings (PC1, PC2) of the principal component analysis on the RDKit Descriptors of our data set and more than 50000 probable odorous molecules (from the GDB-17 database \cite{ruddigkeit_2012} downsampled molecules with a odor probability of 50\% or more which was predicted with the best-performing model according to \cite{mayhew_2022}).}
    \centering
    \begin{tabular}{lrrr}
    \hline
    Feature PC1 & PC1 Loading & Feature PC2 & PC2 Loading \\
    \hline
    Chi1v & 0.170900 & NumHeteroatoms & 0.173622 \\
    Chi1n & 0.170844 & NOCount & 0.166007 \\
    Chi2v & 0.169255 & MinAbsPartialCharge & 0.164090 \\
    Chi2n & 0.168528 & MaxPartialCharge & 0.163948 \\
    NumValenceElectrons & 0.168341 & HallKierAlpha & -0.163185 \\
    HeavyAtomCount & 0.167844 & TPSA & 0.159375 \\
    Chi1 & 0.166394 & SMR\_VSA10 & 0.151925 \\
    LabuteASA & 0.166337 & NumHAcceptors & 0.147437 \\
    Chi0n & 0.163871 & SPS & -0.144280 \\
    Chi0v & 0.163191 & NumAliphaticRings & -0.136433 \\
    ExactMolWt & 0.162538 & NumSaturatedRings & -0.135090 \\
    MolWt & 0.162495 & NumAliphaticCarbocycles & -0.129865 \\
    Chi3v & 0.161745 & MaxAbsEStateIndex & 0.128348 \\
    Chi3n & 0.159376 & MaxEStateIndex & 0.128348 \\
    HeavyAtomMolWt & 0.159341 & NumSaturatedCarbocycles & -0.123744 \\
    \hline
    \end{tabular}
    \label{tab:pca_loadings_w_background}
\end{table}

\clearpage

\section{Hyperparameter ranges}

Here, we provide the type, options, sampling and further conditions of the tuned hyperparameters for each molecular encoder and predictor. Hyperparameters were optimized using the tree-structured Parzen estimator (TPE) via Optuna \cite{akiba_2019}.

\subsection{Encoders}

The tuned hyperparameters of the Morgan fingerprint\cite{morgan_1965, rogers_2010} (\autoref{tab:hp_morgan}), the RDKit fingerprint\cite{RdkitRdkit2025_03_5} (\autoref{tab:hp_rdkit_fp}), the topological torsion fingerprint\cite{nilakantan_1987} (\autoref{tab:hp_topological_fp}), the atom pair fingerprint\cite{carhart_1985} (\autoref{tab:hp_atom_pair_fp}) and ChemBERTa-2\cite{ahmad_2022} (\autoref{tab:hp_chemberta}) are provided.
No hyperparameters of the MACCS-keys fingerprint\cite{durant_2002} and the RDKit descriptors\cite{RdkitRdkit2025_03_5} were tuned.

\begin{table}[h]
\small
\centering
\caption{\ Morgan fingerprint optuna hyperparameter configuration.}
\begin{tabular}{p{4cm}p{2.6cm}p{3cm}p{6cm}}
\hline
Hyperparameter& Type & Options & Sampling\\
\hline
radius & integer & 2, 3, 4&  uniform \\
count & categorical (bool) & True, False&  uniform \\
countSimulation & categorical (bool) & True, False&  uniform \\
fpSize & integer & 1024, 1536, 2048&  uniform\\
includeChirality & categorical (bool) & True, False&  uniform \\
useBondTypes & categorical (bool) & True, False&  uniform \\
atomInvariantsGenerator & categorical & True, None&  uniform \\
\hline
\end{tabular}
\label{tab:hp_morgan}
\end{table}

\begin{table}[h]
\small
\centering
\caption{\ RDKit fingerprint optuna hyperparameter configuration.}
\begin{tabular}{p{4cm}p{2.6cm}p{3cm}p{6cm}}
\hline
Hyperparameter & Type & Options & Sampling\\
\hline
count & categorical (bool) & True, False&  uniform \\
countSimulation & categorical (bool) & True, False&  uniform \\
fpSize & integer & 1024, 1536, 2048&  uniform\\
atomInvariantsGenerator & categorical & True, None&  uniform \\
minPath & integer & 1, 2, 3&  uniform \\
maxPath & integer & 5, 6, 7, 8, 9&  uniform \\
useHs & categorical (bool) & True, False&  uniform \\
branchedPaths & categorical (bool) & True, False&  uniform \\
useBondOrder & categorical (bool) & True, False&  uniform \\
numBitsPerFeature & integer & 1, 2, 3&  uniform \\
\hline
\end{tabular}
\label{tab:hp_rdkit_fp}
\end{table}

\begin{table}[h]
\small
\centering
\caption{\ Topological torsion fingerprint optuna hyperparameter configuration.}
\begin{tabular}{p{4cm}p{2.6cm}p{3cm}p{6cm}}
\hline
Hyperparameter & Type & Options & Sampling\\
\hline
count & categorical (bool) & True, False&  uniform \\
countSimulation & categorical (bool) & True, False&  uniform \\
fpSize & integer & 1024, 1536, 2048&  uniform\\
includeChirality & categorical (bool) & True, False&  uniform \\
torsionAtomCount & integer & 2, 3, 4, 5, 6&  uniform \\
\hline
\end{tabular}
\label{tab:hp_topological_fp}
\end{table}

\begin{table}[h]
\small
\centering
\caption{\ Atom pair fingerprint optuna hyperparameter configuration.}
\begin{tabular}{p{4cm}p{2.6cm}p{3cm}p{6cm}}
\hline
Hyperparameter & Type & Options & Sampling\\
\hline
count & categorical (bool) & True, False&  uniform \\
countSimulation & categorical (bool) & True, False&  uniform \\
minDistance & integer & 1, 2, 3&  uniform \\
maxDistance & integer & 5, 6, 7, 8, 9&  uniform \\
includeChirality & categorical (bool) & True, False&  uniform \\
fpSize & integer & 1024, 1536, 2048&  uniform \\
atomInvariantsGenerator & categorical & True, None&  uniform \\
\hline
\end{tabular}
\label{tab:hp_atom_pair_fp}
\end{table}

\begin{table}[h]
\small
\centering
\caption{\ ChemBERTa optuna hyperparameter configuration.}
\begin{tabular}{p{4cm}p{2.6cm}p{3cm}p{6cm}}
\hline
Hyperparameter & Type & Options & Sampling\\
\hline
target\_layer & integer & 0, 1, 2, 3, 4, 5, 6 &  uniform \\
pooling & categorical & 'mean', 'cls'&  uniform \\
\hline
\end{tabular}
\label{tab:hp_chemberta}
\end{table}

\subsection{Predictors}

The tuned hyperparameters of the predictors logistic regression\cite{verhulst_1845}(\autoref{tab:hp_logistic_regression}), random forest\cite{breiman_2001} (\autoref{tab:hp_rf}), extreme gradient boosting (XGBoost)\cite{chen_2016} (\autoref{tab:hp_xgb}), multi-layer-perceptron (MLP)\cite{rumelhart_1986, adam_2019} and consistent rank logits (CORAL)\cite{cao_2020} (\autoref{tab:hp_mlp}) are provided. In addition, the message-passing neural network framework ChemProp\cite{yang_2019, heid_2024} both with and without initialization from the foundational model CheMeleon\cite{burns_2025} was optimized. The tested hyperparameters are in \autoref{tab:hp_chemprop}.
No hyperparameters of the average prediction were tuned

\begin{table}[h]
\small
\centering
\caption{\ Logistic regression optuna hyperparameter configuration.}
\begin{tabular}{p{4cm}p{2cm}p{3cm}p{2cm}p{4cm}}
\hline
Hyperparameter & Type & Options & Sampling& Condition \\
\hline
standardizer\_name & categorical & 'standard', 'robust', 'minmax', 'yeo-johnson', None&  uniform & — \\
penalty & categorical & 'l2', 'l1', 'elasticnet', None&  uniform & — \\
tol & float & 1e-5-1e-3& log-uniform& — \\
C & float & 1e-2-1e2& log-uniform& defined if penalty not None \\
l1\_ratio & float & 0.1-0.9& uniform (step size 0.1)& defined if penalty = 'elasticnet' \\
solver & categorical & 'liblinear', 'saga'&  uniform & if penalty = 'l1' \\
solver & categorical & 'saga', 'newton-cg', 'lbfgs'&  uniform & if penalty = None \\
solver & categorical & 'liblinear', 'saga', 'newton-cg', 'lbfgs'&  uniform & if penalty = 'l2' \\
binarize\_labels & categorical (bool) & True, False&  uniform & if labels ordinal (not only binary)\\
dealing\_with\_incosistency & categorical & 'sum', 'max'&  uniform & defined if binarize\_labels = True \\
\hline
\end{tabular}
\label{tab:hp_logistic_regression}
\end{table}

\begin{table}[h]
\small
\centering
\caption{\ Random forest optuna hyperparameter configuration.}
\begin{tabular}{p{4cm}p{2cm}p{3cm}p{2cm}p{4cm}}
\hline
Hyperparameter & Type & Options & Sampling& Condition \\
\hline
objective & categorical & 'classification', 'regression'&  uniform & — \\
n\_estimators & integer & 50-500&  uniform (step size 50) & — \\
max\_depth & integer & 5-20&  uniform (step size 5) & — \\
min\_samples\_split & integer & 2, 3, 4&  uniform & — \\
min\_samples\_leaf & integer & 1, 3, 5&  uniform & — \\
min\_weight\_fraction\_leaf & float & 0.0, 0.2, 0.4&  uniform & — \\
max\_features & categorical & 'sqrt', 'log2', None&  uniform & — \\
max\_leaf\_nodes & integer & 25, 50, 75, 100&  uniform & — \\
min\_impurity\_decrease & float & 0.0, 0.05, 0.1&  uniform & — \\
criterion & categorical & 'gini', 'entropy', 'log\_loss'&  uniform & if objective = 'classification'\\
binarize\_labels & categorical (bool) & True, False&  uniform& if objective = 'classification' and labels ordinary (not only binary)\\
dealing\_with\_incosistency & categorical & 'sum', 'max'&  uniform & if binarize\_labels = True\\
\hline
\end{tabular}
\label{tab:hp_rf}
\end{table}

\begin{table}[h]
\small
\centering
\caption{\ XGBoost Hyperparameter Space with Conditional Dependencies}
\begin{tabular}{p{4cm}p{2cm}p{3cm}p{2cm}p{4cm}}
\hline
Hyperparameter & Type & Options& Sampling& Condition \\
\hline
objective & categorical & 'binary:logistic', 'multi:softmax', 'reg:squarederror'&  uniform& only binary labels: always 'binary:logistic'\\
custom\_metric & fixed & 'mse\_macro' or 'f1\_score' & — & only binary labels: 'f1\_score'; ordinary labels: 'mse\_macro'\\
booster & categorical & 'gbtree', 'dart'&  uniform & — \\
num\_boost\_round & fixed & 250 & — & — \\
early\_stopping\_rounds & fixed & 50 & — & — \\
eta & float & 0.005-0.5& log-uniform & — \\
lambda & float & 0.001-10& log-uniform & — \\
alpha & float & 0.001-10& log-uniform & — \\
gamma & float & 0.001-10& log-uniform & — 
\\
max\_depth & integer & 4, 6, 8, 10&  uniform & — 
\\
min\_child\_weight & float & 0.01-100& log-uniform & — 
\\
max\_delta\_step & float & 0, 1, 2&  uniform & — 
\\
subsample & float & 0.6, 0.8, 1.0&  uniform & — 
\\
colsample\_bytree & float & 0.6, 0.8, 1.0&  uniform & — 
\\
colsample\_bylevel & float & 0.6, 0.8, 1.0&  uniform & — 
\\
num\_parallel\_tree & integer & 1, 2, 3&  uniform & — \\
rate\_drop & float & 0.0, 0.2, 0.4&  uniform & booster = 'dart' \\
skip\_drop & float & 0.0, 0.2, 0.4&  uniform & booster = 'dart' \\
normalize\_type & categorical & 'tree', 'forest'&  uniform & booster = 'dart' \\
sample\_type & categorical & 'uniform', 'weighted'&  uniform & booster = 'dart' \\
binarize\_labels & categorical (bool) & True, False&  uniform& if labels ordinal (not only binary)\\
dealing\_with\_incosistency & categorical & 'sum', 'max'&  uniform & if binarize\_labels = True \\
multi\_strategy & categorical & 'one\_output\_per\_tree', 'multi\_output\_tree'&  uniform & if binarize\_labels = True and booster = 'gbtree' \\
num\_class & fixed & 4 & — & if objective = 'multi:softmax' \\
\hline
\end{tabular}
\label{tab:hp_xgb}
\end{table}

\begin{table}[h]
\small
\centering
\caption{\ MLP and CORAL predictor optuna hyperparameter configuration.}
\begin{tabular}{p{4cm}p{2cm}p{3cm}p{2cm}p{4cm}}
\hline
Hyperparameter & Type & Options& Sampling& Condition \\
\hline
standardizer\_name & categorical & 'standard', 'robust', 'minmax', 'yeo-johnson', None&  uniform & — \\
n\_layers & integer & 2-12& uniform (step size 2) & — \\
dim & integer & 32-512& log-uniform & — \\
batch\_size & categorical & 16, 32, 64, 128, 256&  uniform & — \\
n\_epochs & fixed & 250 & — & — \\
early\_stopping\_rounds & fixed & 50 & — & — \\
delta & fixed & 0 & — & — \\
objective\_name & categorical & 'binary\_crossentropy', 'mse', 'coral'& uniform & 'coral' only for CORAL predictor (fixed); if labels binary only 'binary\_crossentropy'\\
learning\_rate & float & 5e-6-5e-4& log-uniform & — \\
optimizer\_name & categorical & 'adam', 'adamw', 'sgd', 'rmsprop', 'adagrad'&  uniform & — \\
training\_scheduler\_name & categorical & None, 'ReduceLROnPlateau', 'CosineAnnealingLR', 'CosineAnnealingWarmRestarts'&  uniform & — \\
warmup\_epochs & integer & 0, 10, 20, 30&  uniform & — \\
activation & categorical & 'relu', 'leaky-relu', 'tanh', 'sigmoid', 'elu'&  uniform & — \\
dropout & float & 0.0-0.5& uniform (step size 0.1)& — \\
weight\_average & categorical & False, 'swa', 'ema'&  uniform & — \\
metric\_name & fixed & 'mse\_macro' or 'f1\_score' & — & binary labels 'f1\_score'; ordinary labels:  'mse\_macro'\\
weight\_decay & float & 1e-5-1e-3& log-uniform & — \\
betas & tuple(float, float) & beta1: 0.85, 0.90, 0.95, beta2: 0.990, 0.993, 0.996, 0.999 & uniform & if optimizer in ['adam', 'adamw'] \\
momentum & float & 0.0-0.5& uniform (step size 0.1)& if optimizer in ['sgd', 'rmsprop'] \\
nesterov & categorical & True, False&  uniform & if optimizer = 'sgd' and momentum > 0 \\
dampening & float & 0.0, 0.05, 0.1& uniform & if optimizer = 'sgd' and nesterov = False \\
alpha & float & 0.9-0.99& uniform (step size 0.03)& if optimizer = 'rmsprop' \\
lr\_decay & float & 0.0, 0.05, 0.1& uniform & if optimizer = 'adagrad' \\
factor & float & 0.1-0.5]& uniform (step size 0.1)& if scheduler = 'ReduceLROnPlateau' \\
T\_0 & integer & 10, 20, 30, 40&  uniform & if scheduler = 'CosineAnnealingWarmRestarts' \\
T\_mult & integer & 2, 3, 4&  uniform & if scheduler = 'CosineAnnealingWarmRestarts' \\
binarize\_labels & categorical (bool) & True, False&  uniform & if labels ordinal (not only binary)\\
dealing\_with\_incosistency & categorical & 'sum', 'max'&  uniform & if binarize\_labels = True \\
\hline
\end{tabular}
\label{tab:hp_mlp}
\end{table}

\begin{table}[h]
\small
\centering
\caption{\ ChemProp and CheMeleon MPNN predictor optuna hyperparameter configuration.}
\begin{tabular}{p{4cm}p{2cm}p{3cm}p{2cm}p{4cm}}
\hline
Hyperparameter & Type & Options& Sampling& Condition \\
\hline
epochs & fixed & 250 ChemProp, 100 CheMeleon& — & — \\
early\_stopping\_rounds & fixed & 50 ChemProp, 25 CheMeleon & — & — \\
delta & fixed & 0 & — & — \\
mode & fixed & 'min' & — & — \\
batch\_size & categorical & 16, 32, 64, 128, 256&  uniform & — \\
aggregation & categorical & 'norm', 'sum', 'mean', 'attentive'&  uniform & — \\
n\_layers & integer & 1–12&  uniform & — \\
objective & categorical & 'binary\_crossentropy', 'mse'&  uniform & if labels binary only 'binary\_crossentropy'\\
predictor\_dim & integer & 32–512&  log-uniform & — \\
predictor\_dropout & float & 0.0–0.5&  uniform (step size 0.1)& — \\
predictor\_activation & categorical & 'relu', 'leakyrelu', 'prelu', 'tanh', 'elu'&  uniform & — \\
warmup\_epochs & integer & 0, 10, 20, 30&  uniform & — \\
initial\_learning\_rate & float & 5e-6–1e-3&  log-uniform & — \\
max\_learning\_rate\_factor & float & 1–100&  log-uniform & used to compute max\_learning\_rate \\
final\_learning\_rate\_factor & float & 0.01–10&  log-uniform & used to compute final\_learning\_rate \\
max\_learning\_rate & derived & — & — & initial\_learning\_rate × max\_learning\_rate\_factor \\
final\_learning\_rate & derived & — & — & initial\_learning\_rate × final\_learning\_rate\_factor \\
message\_passing & categorical & 'bond', 'atom', 'chemeleon'& uniform & 'chemeleon' only for CheMeleon predictor (fixed)\\
mp\_dim & integer & 32–512&  log-uniform & only ChemProp predictor\\
bias & categorical (bool) & True, False&  uniform & only ChemProp predictor\\
n\_message\_passing\_iterations & integer & 1–8&  uniform & only ChemProp predictor\\
messages\_undirected\_edges & categorical (bool) & True, False&  uniform & only ChemProp predictor\\
mp\_dropout & float & 0.0–0.5&  uniform (step size 0.1)& only ChemProp predictor\\
mp\_activation & categorical & 'relu', 'leakyrelu', 'prelu', 'tanh', 'elu'&  uniform & only ChemProp predictor\\
binarize\_labels & categorical (bool) & True, False&  uniform & if labels ordinal (not only binary)\\
dealing\_with\_incosistency & categorical & 'sum', 'max'&  uniform & if binarize\_labels = True \\
\hline
\end{tabular}
\label{tab:hp_chemprop}
\end{table}

\clearpage

\section{Direct and indirect approach model performance}

Regarding the direct approach, using the average of labels achieved a macro MSE of 1.5, while the message passing neural network ChemProp \cite{yang_2019, heid_2024} achieved 0.65, and the foundational model CheMeleon \cite{burns_2025} 0.54.

\begin{figure}[h]
    \centering
    \includegraphics[scale=1]{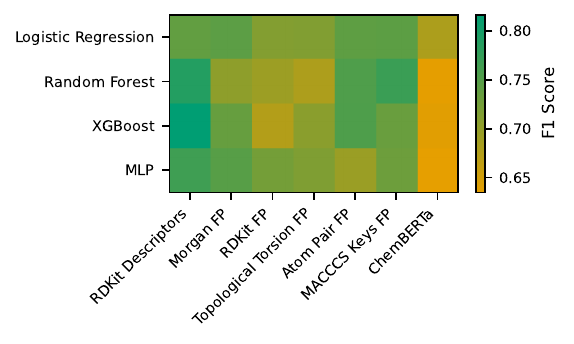}
    \caption{Model performance first step of the indirect approach (binary classifier). F1-score (target minority class) of the hold-out test set of all combinations of molecule encoder (bottom) and predictor (left). MLP is multi-layer-perceptron and FP fingerprint. The message passing neural network ChemProp \cite{yang_2019, heid_2024} achieved an F1-score of 0.77, and the foundational model CheMeleon \cite{burns_2025} of 0.74.}
    \label{fig:heatmap_has_odor}
\end{figure}

\begin{figure}[h]
    \centering
    \includegraphics[scale=1]{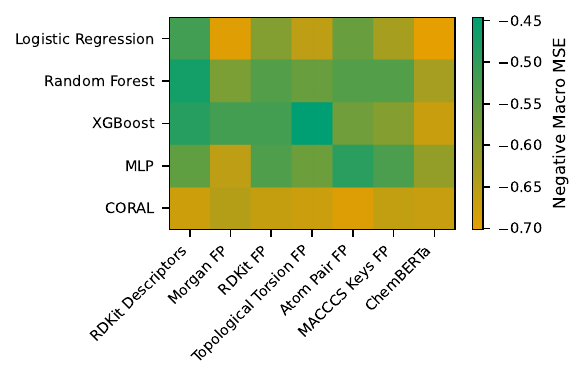}
    \caption{Model performance second step of the indirect approach. Macro averaged mean squared error (MSE) over the categories of the hold-out test set of all combinations of molecule encoder (bottom) and predictor (left). MLP is multi-layer-perceptron and FP fingerprint. The message passing neural network using ChemProp \cite{yang_2019, heid_2024} achieved a score of 0.56, and the foundational model CheMeleon \cite{burns_2025} of 0.51.}
    \label{fig:heatmap_wo_odorless}
\end{figure}

\begin{figure}[h]
    \centering
    \includegraphics[scale=1]{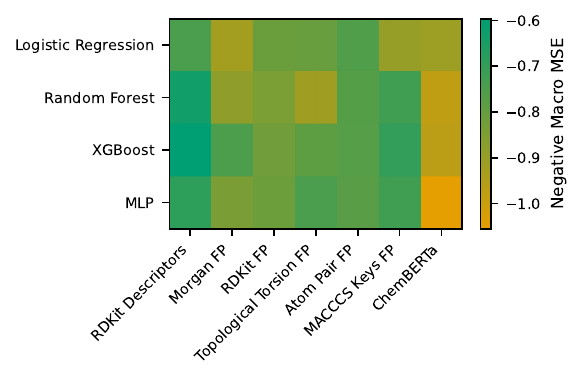}
    \caption{Model performance combined steps of the indirect approach. Macro averaged mean squared error (MSE) over the categories of the hold-out test set of the cross-validation for the best hyperparameter optimized models of the same model combinations between the steps (molecule encoder (bottom) and predictor (left)). MLP is multi-layer-perceptron and FP fingerprint. The message passing neural network using ChemProp \cite{yang_2019, heid_2024} achieved a score of 0.70, and the foundational model CheMeleon \cite{burns_2025} of 0.79.}
    \label{fig:heatmap_two_steps}
\end{figure}

\begin{figure}[h]
    \centering
    \includegraphics[scale=1]{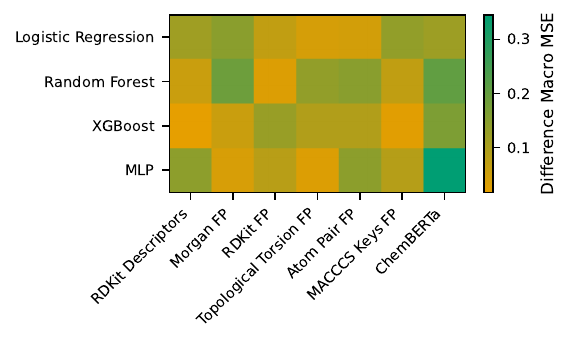}
    \caption{Differences in macro MSE between the indirect and direct approach models on the hold-out test set (direct subtracted from indirect). For each combination of molecular encoder (bottom) and predictor (left), the macro MSE of the direct model was subtracted from the indirect model of the same model combinations. MLP is multi-layer-perceptron and FP fingerprint. The difference regarding the message passing neural network using ChemProp \cite{yang_2019, heid_2024} was 0.05, and the foundational model CheMeleon \cite{burns_2025} 0.25.}
    \label{fig:heatmap_difference}
\end{figure}

Results of odor strength prediction using the indirect approach, consisting of two models to predict first whether a molecule is odorous and if so to predict its odor strength, are further shown in this section. \autoref{fig:heatmap_has_odor}, \autoref{fig:heatmap_wo_odorless} and \autoref{fig:heatmap_two_steps} provide a comparison of the first, second and combined steps, respectively, across the tested algorithms for molecular encoding and odor prediction. The macro MSE difference to the direct approach is shown in \autoref{fig:heatmap_difference}.

\begin{figure}[h]
    \centering
    \includegraphics[scale=1]{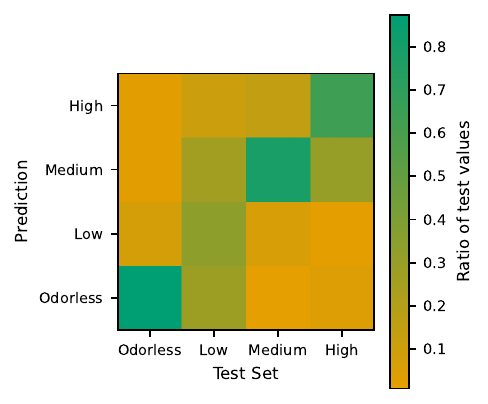}
    \caption{Model performance combined steps of the indirect approach. Confusion matrix normed by the number of the test values of the combined best-performing models (RDKit descriptors with XGBoost and topological torsion fingerprint with XGBoost) on the test set for 10 random seeded training runs (MSE macro: 0.59, R\textsuperscript{2}: 0.53).}
\label{fig:model_performance_indirect}
\end{figure}

\begin{figure}[h]
    \centering
    \includegraphics[scale=1]{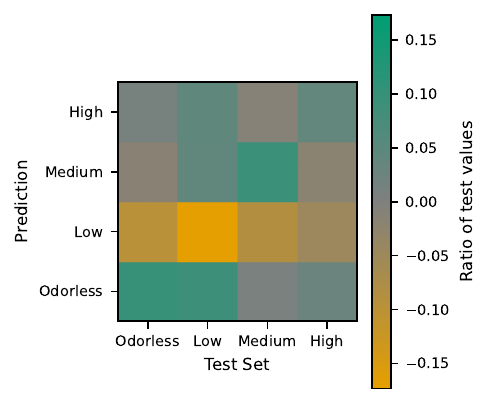}
    \caption{Differences in the model performance of indirect and direct approach (direct subtracted from indirect). The confusion matrix normed by the number of the test values of the combined best-performing model of the direct approach was subtracted from the one of the indirect approach. The matrices were calculated on the test set for 10 random seeded training runs.}
\label{fig:model_performance_difference}
\end{figure}

The best results of the indirect approach achieved the RDKit descriptors with XGBoost for the first step and the topological torsion fingerprint with XGBoost for the second step.
The performance on a hold-out test set was only slightly and not significantly worse than the best-performing model of the direct approach. The confusion matrix normed by the number of each odor strength category in the test of the best-performing indirect approach model set is shown in \autoref{fig:model_performance_indirect} and the difference to the confusion matrix of the direct approach in \autoref{fig:model_performance_difference}. The indirect model performed slightly better regarding odorless, medium and high odor strength but it predicted less low odor strengths and performed significantly poorer in this category. In addition, there were marginally more severe errors (odorless-high, high-odorless).

\clearpage

\section{Best-performing Model Validation}

\begin{figure*}[!h]
    \centering
    \includegraphics[scale=1]{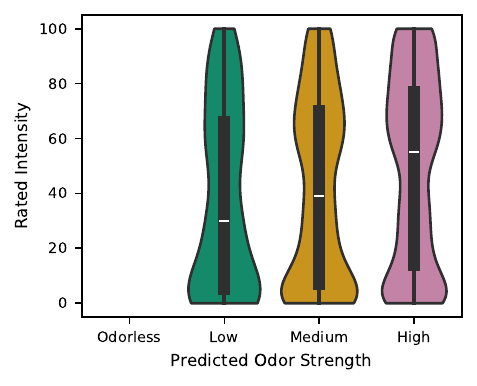}
    \caption{Area normed violin plots of the best-performing model predictions of the novel molecules from Keller \textit{et al.}\cite{keller_2017} and their rated intensities (from 0 to 100) at 10\textsuperscript{-5} dilution. No molecules were predicted as odorless, which is consistent with the expectation that nearly odorless compounds are unlikely to be used at such low concentrations}
    \label{fig:violin_1e-05}
\end{figure*}

Here, we provide \autoref{fig:violin_1e-05}, which shows normed violin plots of experimental intensity ratings in comparison to our best-performing model predictions (direct approach) of the same molecules. It demonstrates the performance of our best-performing model on an independent hold-out test set of intensity ratings from Keller \textit{et al.}\cite{keller_2017} at another dilution level of 1/10\textsuperscript{5}. The trend is similar to the lower dilution level in the main part: an increase in predicted odor strength correlates with a higher odor intensity rating.

\clearpage

\section{SHAP Feature Importance Analysis}

\begin{figure}[!h]
    \centering
    \includegraphics[scale=1]{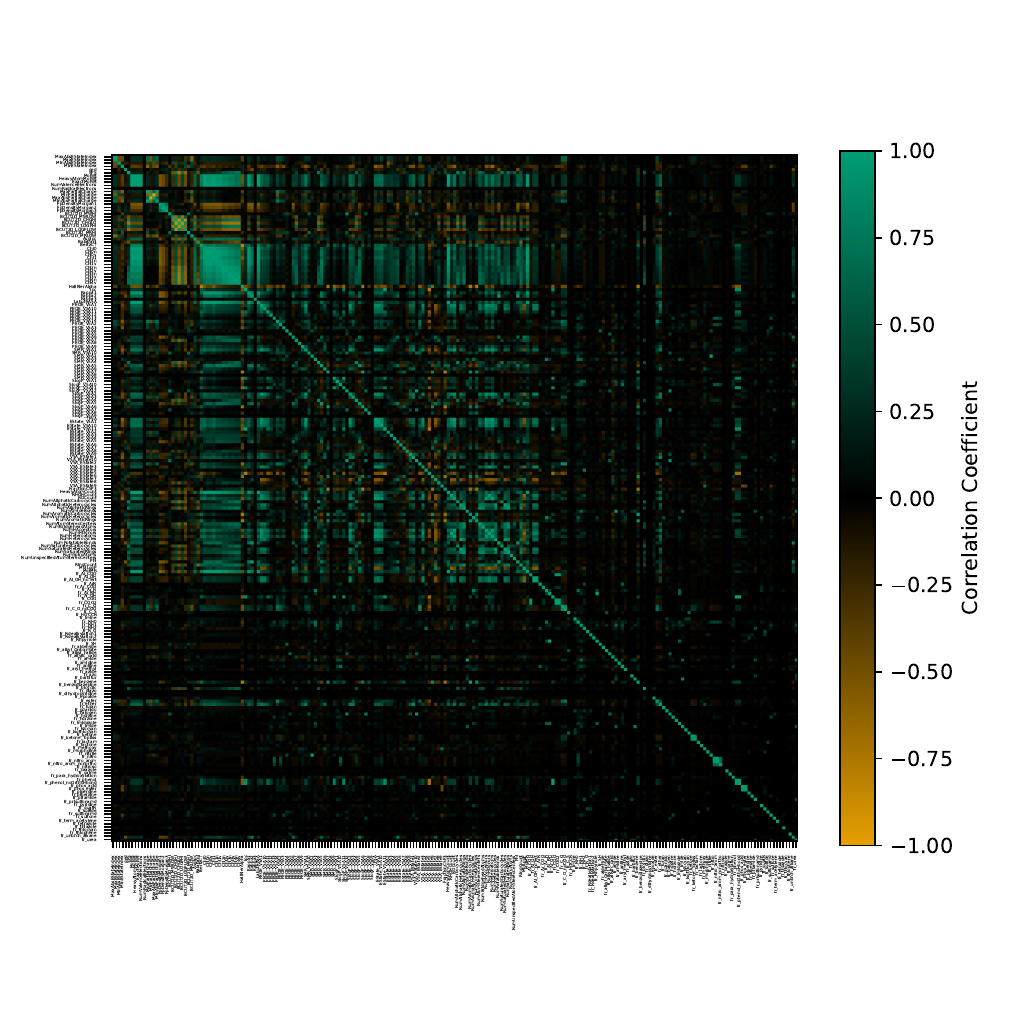}
    \caption{Correlation matrix of the 217 RDKit Descriptors of the molecules of our data set.}
    \label{fig:correlation_matrix}
\end{figure}

\begin{figure}[!h]
    \centering
    \includegraphics[scale=1]{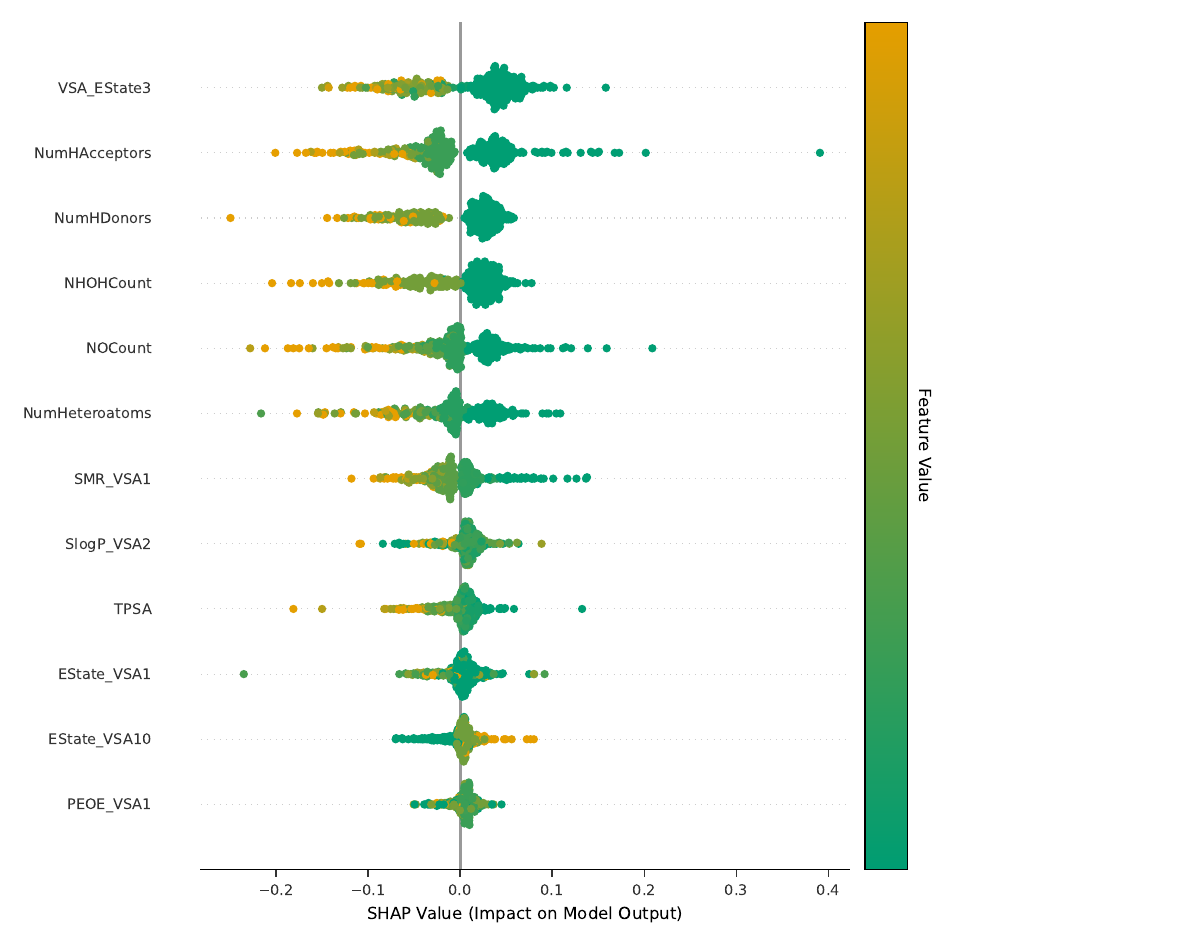}
    \caption{SHAP feature importance of the features of the 'Polarity' group colored by the relative feature value on the test set using the train set as background. Each dot is an instance of the test set.}
    \label{fig:polarity}
\end{figure}

\begin{figure}[!h]
    \centering
    \includegraphics[scale=1]{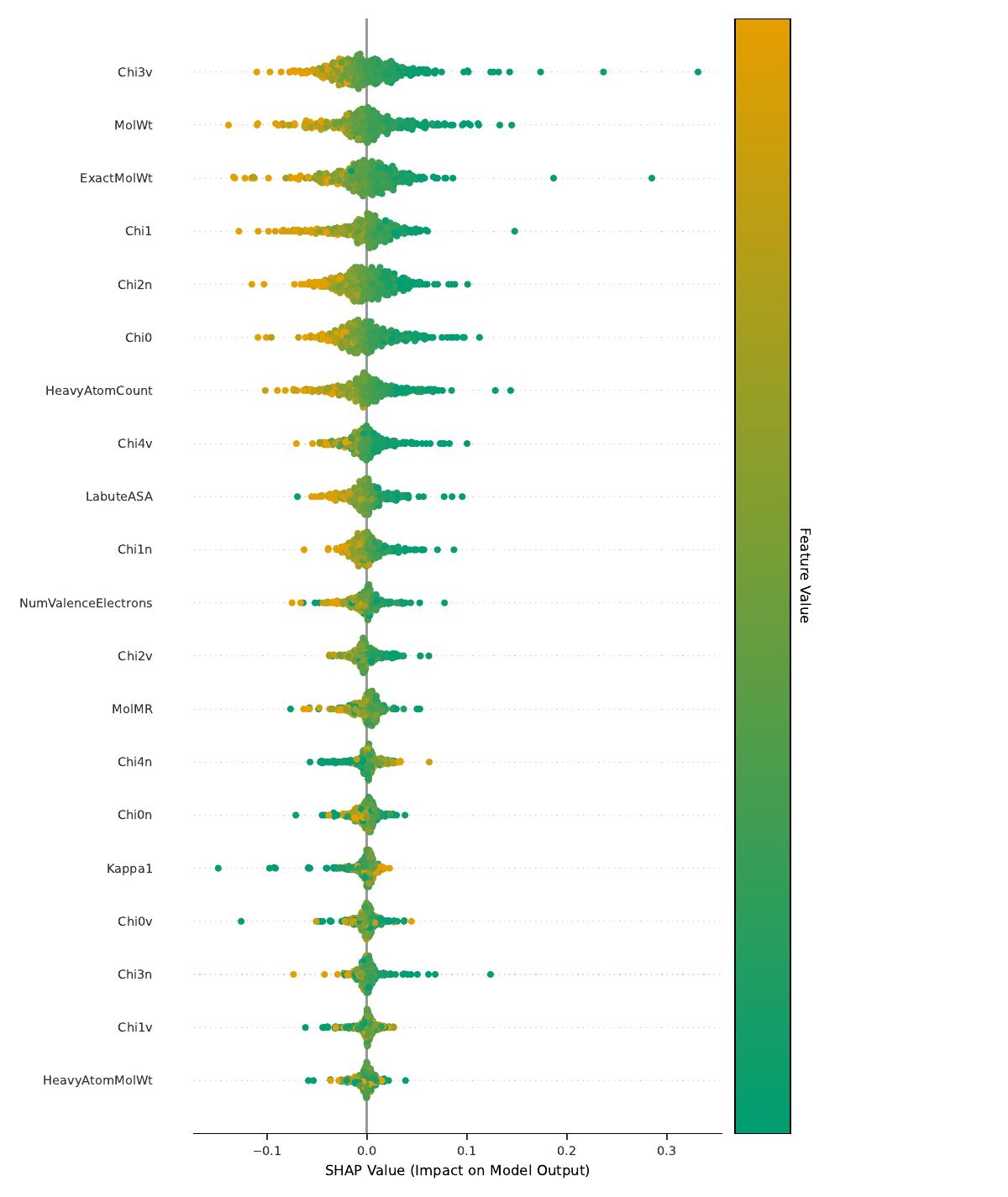}
    \caption{SHAP feature importance of the features of the 'Weight and Shape' group colored by the relative feature value on the test set using the train set as background. Each dot is an instance of the test set.}
    \label{fig:weight_and_shape}
\end{figure}

\begin{figure}[!h]
    \centering
    \includegraphics[scale=1]{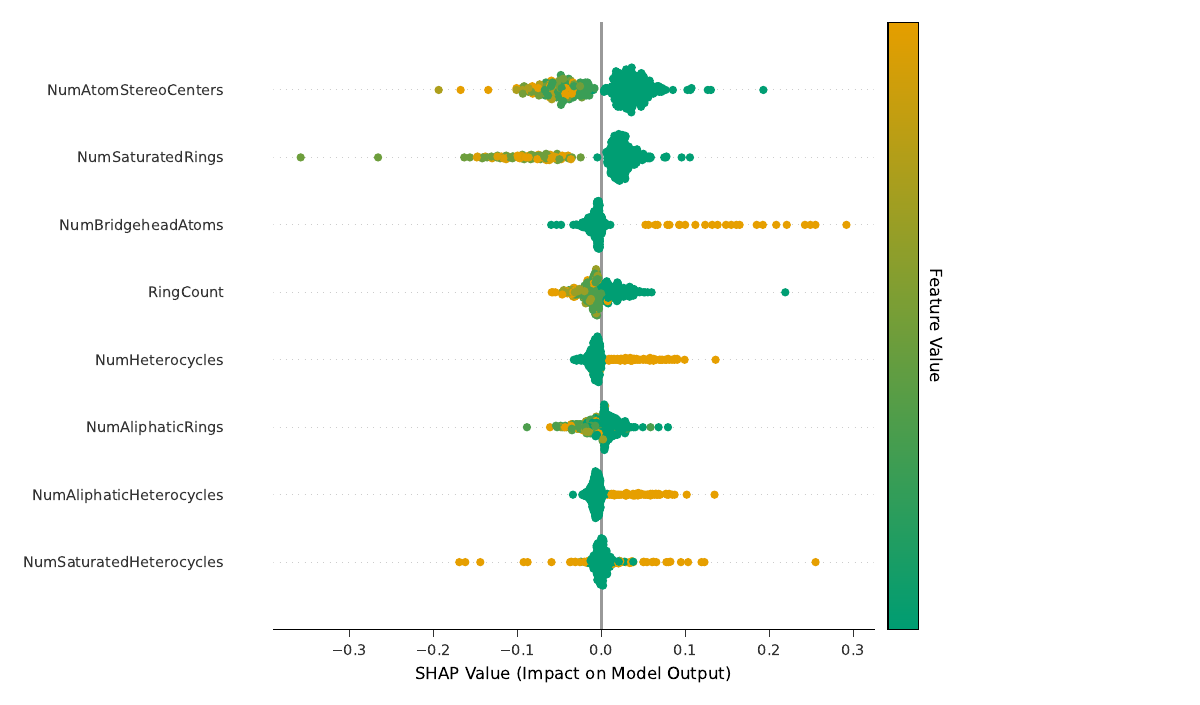}
    \caption{SHAP feature importance of the features of the 'Rings' group colored by the relative feature value on the test set using the train set as background. Each dot is an instance of the test set.}
    \label{fig:rings}
\end{figure}

\begin{figure}[!h]
    \centering
    \includegraphics[scale=1]{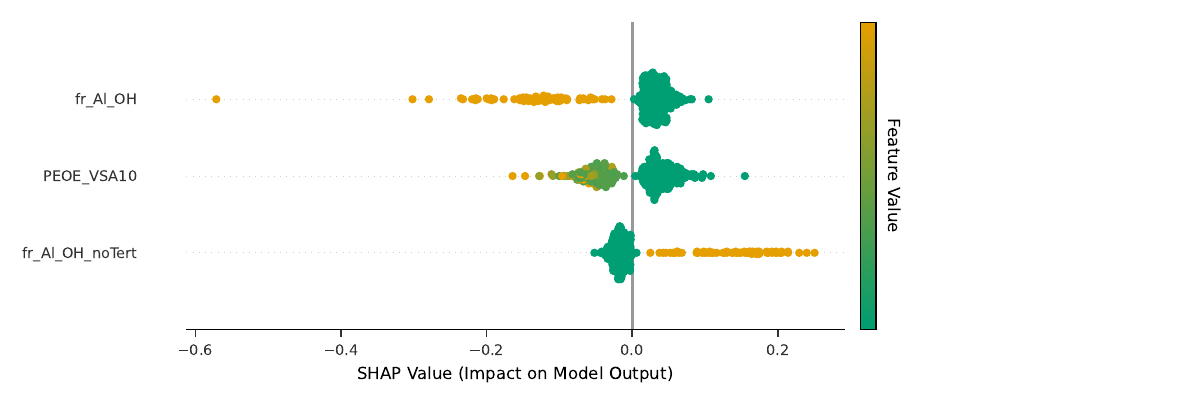}
    \caption{SHAP feature importance of the features of the 'Alcohol Groups' group colored by the relative feature value on the test set using the train set as background. Each dot is an instance of the test set.}
    \label{fig:alcohol_groups}
\end{figure}

\begin{figure}[!h]
    \centering
    \includegraphics[scale=1]{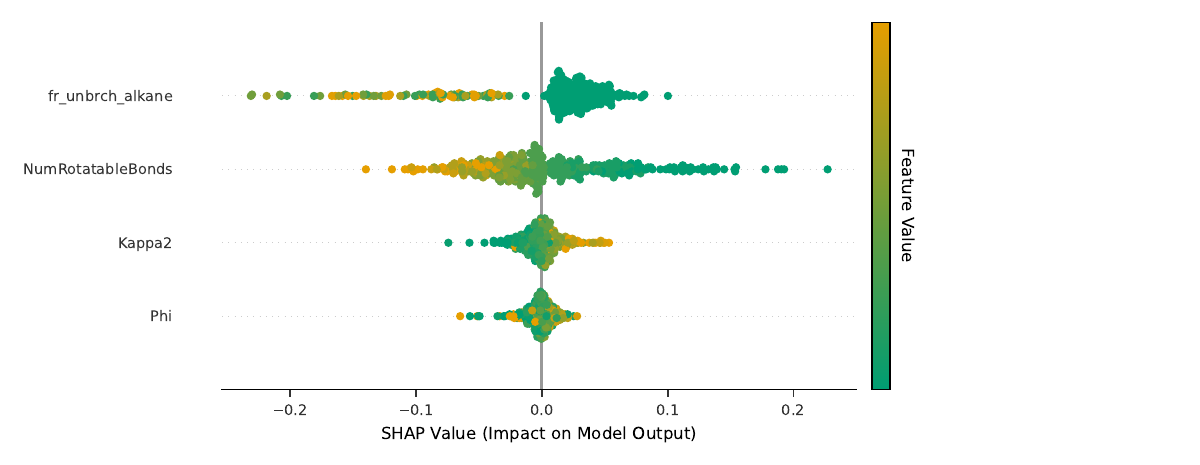}
    \caption{SHAP feature importance of the features of the 'Branching' group colored by the relative feature value on the test set using the train set as background. Each dot is an instance of the test set.}
    \label{fig:branching}
\end{figure}

\begin{figure}[!h]
    \centering
    \includegraphics[scale=1]{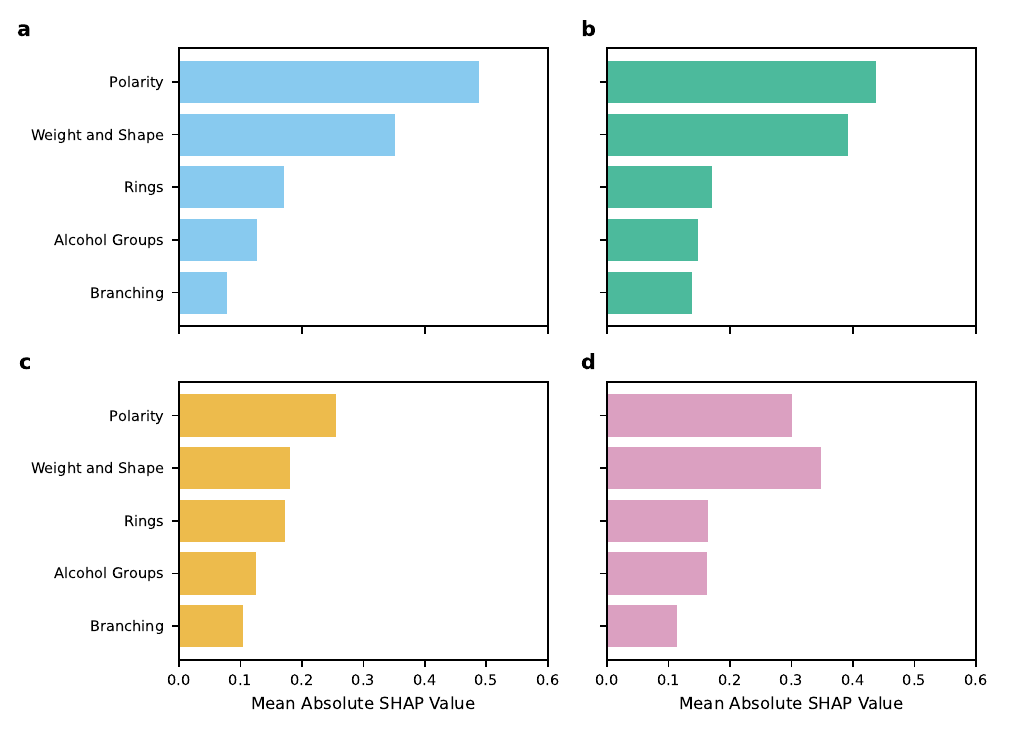}
    \caption{SHAP feature importance per odor strength prediction category: (a) odorless, (b) low, (c) medium and (d) high odor strength.}
    \label{fig:shap_by_odor_strength}
\end{figure}

We wanted to investigate feature importance of our best-performing model (direct approach: RDKit Descriptors with MLP). However, we identified several correlated features as shown in \autoref{fig:correlation_matrix}. Consequently, we grouped highly correlated features (correlation threshold 0.75) via agglomerative clustering. The SHAP (SHapley Additive exPlanations) values on the test set (train set as background) of each feature of the five most influential groups are shown in \autoref{fig:polarity} (Polarity), \autoref{fig:weight_and_shape} (Weight and Shape), \autoref{fig:rings} (Rings), \autoref{fig:alcohol_groups} (Alcohol Groups) and \autoref{fig:branching} (Branching). \autoref{fig:shap_by_odor_strength} shows the SHAP feature importance of these groups per predicted odor strengths.

\begin{figure}[!h]
    \centering
    \includegraphics[scale=1]{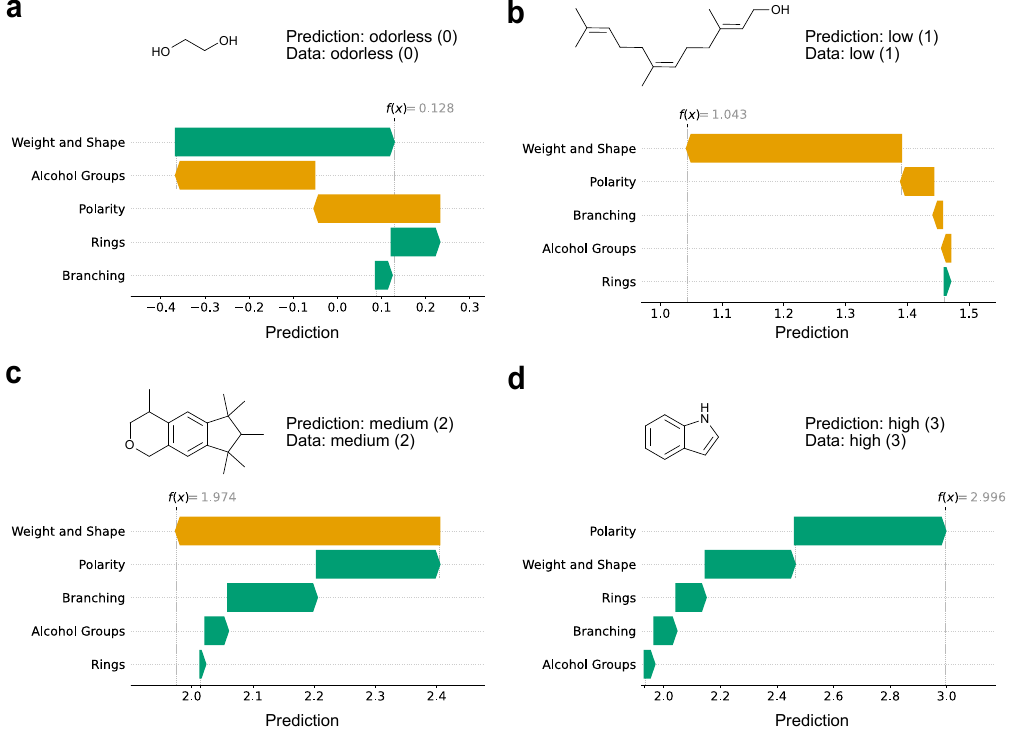}
    \caption{SHAP waterfall plots of the global most important feature groups for 4 example molecules of the data set using the train set as background. The contribution of each feature (SHAP value) to the prediction to deviate from the average background is shown. The molecules are (a) Ethylene glycol (odorless), (b) \textit{E,Z}-Farnesol (low odor strength), (c) 1,3,4,6,7,8-Hexahydro-4,6,6,7,8,8-hexamethylcyclopenta[g]-2-benzopyran, also known as Galaxolide (medium odor strength), (d) Indole (high odor strength).}
    \label{fig:shap_local}
\end{figure}

\clearpage
Local SHAP feature group contributions for four representative molecules are shown in \autoref{fig:shap_local} to illustrate local interpretations. Smaller molecules, such as Ethylene Glycol or Indole, tend to shift the prediction toward higher odor strength, whereas larger molecules, such as \textit{E,Z}-Farnesol or Galaxolide, shift it toward lower odor strength. Furthermore, the prediction for the highly polar ethylene glycol is driven by the negative contributions of its alcohol groups and overall polarity, while the lower but sufficient polarity of Galaxolide and Indole contributes positively to their higher odor strengths.

\clearpage

\section{Further Method Details}

The following keywords were used to map a PubChem description to a corresponding odor strength:
\begin{itemize}
    \item \textbf{Odor strength: description}
    \item odorless: odourless, odorless, no odour, no odor, very faint, very weak, very mild
    \item low: faint, weak
    \item high: very strong, very intense, very powerful, very pungent, very aromatic
\end{itemize}

The following 13 molecules in \autoref{tab:very_high_odor_strengths} were originally labeled as 'very high' odor strength and reclassified to 'high' odor strength.

\begin{table}[h]
\centering
\begin{tabular}{lll}
\hline
 Name & CAS & Canonical SMILES \\
\hline
ortho-thioguaiacol & 7217-59-6 & COc1ccccc1S \\
(Z)-6-nonenal & 2277-19-2 & CC/C=C\textbackslash CCCCC=O \\
skatole & 83-34-1 & Cc1c[nH]c2ccccc12 \\
isopropyl mercaptan & 75-33-2 & CC(C)S \\
caramel furanone & 28664-35-9 & CC1=C(O)C(=O)OC1C \\
2-mercaptopropionic acid & 79-42-5 & CC(S)C(=O)O \\
cortex pyridine & 2110-18-1 & c1ccc(CCCc2ccccn2)cc1 \\
(Z)-4-heptenal & 6728-31-0 & CC/C=C\textbackslash CCC=O \\
2-acetyl pyrazine & 22047-25-2 & CC(=O)c1cnccn1 \\
2-acetyl thiazole & 24295-03-2 & CC(=O)c1nccs1 \\
propyl mercaptan & 107-03-9 & CCCS \\
maple furanone & 698-10-2 & CCC1OC(=O)C(O)=C1C \\
Allyl Isothiocyanate &  & C=CCN=C=S \\
\hline
\end{tabular}
    \caption{Molecules which were originally labeled as 'very high' odor strength and reclassified to 'high' odor strength.}
    \label{tab:very_high_odor_strengths}
\end{table}

\providecommand*{\mcitethebibliography}{\thebibliography}
\csname @ifundefined\endcsname{endmcitethebibliography}
{\let\endmcitethebibliography\endthebibliography}{}